\documentclass[useAMS,twocolumn,usenatbib]{mn2e}
\usepackage{times,psfig,epsfig} 
\usepackage{rotating, graphicx,dsfont}
\usepackage{color}
\usepackage{amssymb, amsmath}

\newcommand{\hm}{\,h^{-1}{\rm Mpc}}
\newcommand{\chandra}{{\it Chandra}}

\newcommand{\rosat}{{\it ROSAT}}
\newcommand{\suzaku}{{\it SUZAKU}}
\newcommand{\xmm}{{\it XMM-Newton}}

\newcommand{\rexcess}{{\it REXCESS}}
\newcommand{\planck}{{\it Planck}}

\newcommand{\Zfe}{\mbox{$Z_{\rmn{Fe}} \,$}}

\newcommand{\Zsife}{\mbox{$Z_{\rmn{Si}}/Z_{\rmn{Fe}}~$}}

\newcommand{\gadget}{{\footnotesize {\sc GADGET~}}}

\newcommand{\nr}{NR}
\newcommand{\w}{CSF}
\newcommand{\agn}{AGN}

\newcommand{\be}{\begin{equation}}
\newcommand{\ee}{\end{equation}}
\newcommand{\ba}{\begin{eqnarray}}
\newcommand{\ea}{\end{eqnarray}}
\newcommand{\brr}{\begin{array}}
\newcommand{\err}{\end{array}}
\newcommand{\bc}{\begin{center}}
\newcommand{\ec}{\end{center}}

\newcommand{\msun}{\,h^{-1}M_\odot}
\newcommand{\hMpc}{\mbox{$h^{-1}{\rmn{Mpc}}~$}}

\newcommand{\vel}{\,{\rm km\,s^{-1}}}

\newcommand{\tsl}{\hbox{$T_\mathrm{sl}$}}

\newcommand{\lxtsl}{\hbox{$L_\mathrm{X}$-$T_{\rm sl}$}}
\newcommand{\stsl}{\hbox{$K$-$T_{\rm sl}$}}

\newcommand{\tslm}{\hbox{$T_{\rm sl}$-$M$}}

\newcommand{\yx}{\hbox{$Y_\mathrm{X}$}}
\newcommand{\yxm}{\hbox{$Y_\mathrm{X}$-$M$}}

\newcommand{\lxtslband}{\hbox{$L_\mathrm{X,[0.1-2.4]\,keV}$-$T_{\rm sl}$}}
\newcommand{\yxmsl}{\hbox{$Y_\mathrm{X,sl}$-$M$}}

\newcommand{\tmw}{\hbox{$T_{\mathrm{mw}}$}}

\newcommand{\stmw}{\hbox{$K$-$T_{\rm mw}$}}

\newcommand{\yxmwm}{\hbox{$Y_\mathrm{X,mw}$-$M$}}
\newcommand{\tmwm}{\hbox{$T_{\rm mw}$-$M$}}

\newcommand{\lxtmwband}{\hbox{$L_\mathrm{X,[0.1-2.4]\,keV}$-$T_{\rm mw}$}}

\newcommand{\mincir}{\raise
  -2.truept\hbox{\rlap{\hbox{$\sim$}}\raise5.truept \hbox{$<$}\ }}
\newcommand{\magcir}{\raise
  -2.truept\hbox{\rlap{\hbox{$\sim$}}\raise5.truept \hbox{$>$}\ }}
\newcommand{\siml}{\raise
  -2.truept\hbox{\rlap{\hbox{$\sim$}}\raise5.truept \hbox{$<$}\ }}
\newcommand{\simg}{\raise
  -2.truept\hbox{\rlap{\hbox{$\sim$}}\raise5.truept \hbox{$>$}\ }}

\def\lesssim{\,\lower2truept\hbox{${<\atop\hbox{\raise4truept\hbox{$\sim$}}}$}\,}
\def\gtrsim{\,\lower2truept\hbox{${>\atop\hbox{\raise4truept\hbox{$\sim$}}}$}\,}

\title[Thermal and chemodynamical ICM properties]
{On the role of AGN feedback on the thermal and chemodynamical properties of the hot intra-cluster medium}

\author[S. Planelles et al.]
{S. Planelles$^{1,2}$\thanks{e-mail: susana.planelles@oats.inaf.it}, S. Borgani$^{1,2,3}$, 
D. Fabjan$^{3,6,7}$,   M. Killedar$^{1,2}$, G. Murante$^{2}$,
\newauthor
 G.~L. Granato$^{2}$, C. Ragone-Figueroa$^{8,2}$, K. Dolag$^{4,5}$ \\~\\
\footnotesize 
$^1$ Astronomy Unit, Department of Physics, University of Trieste, via Tiepolo 11, I-34131 Trieste, Italy\\
$^2$ INAF, Osservatorio Astronomico di Trieste, via Tiepolo 11, I-34131 Trieste, Italy\\ 
$^3$ INFN -- National Institute for Nuclear Physics, Via Valerio 2, I-34127 Trieste, Italy\\ 
$^4$ Universit\"atssternwarte M\"unchen, Scheinerstr. 1, D-81679 M\"unchen, Germany\\
$^5$ Max-Planck-Institut f\"ur Astrophysik, P.O. Box 1317, D-85741 Garching, Germany\\
$^6$ SPACE-SI, Slovenian Centre of Excellence for Space 
          Sciences and Technologies, A$\check{s}$ker$\check{c}$eva 12, 1000 Ljubljana, Slovenia \\
$^7$ Faculty of Mathematics and Physics, University of Ljubljana, Jadranska 19, 1000 Ljubljana, Slovenia \\
$^8$ Instituto de Astronom\'ia Te\'orica y Experimental (IATE),\\
  Consejo Nacional de Investigaciones Cient\'ificas y T\'ecnicas de la Rep\'ublica Argentina (CONICET),\\ Observatorio
  Astron\'omico, Universidad Nacional de C\'ordoba, Laprida 854, X5000BGR, C\'ordoba, Argentina\\
 }

\begin{document}
\maketitle 

\begin{abstract}
 We present an analysis of the properties of the intra-cluster medium
 (ICM) in an extended set of cosmological hydrodynamical simulations of
 galaxy clusters and groups performed with the TreePM+SPH {\tt
 GADGET-3} code.  Besides a set of non-radiative simulations, we
  carried out two sets of simulations including radiative cooling,
  star formation, metal enrichment and feedback from supernovae (SNe),
  one of which also accounts for the effect of feedback from active
  galactic nuclei (AGN) resulting from gas accretion onto
  super-massive black holes. These simulations are analysed with the
  aim of studying the relative role played by SN and AGN feedback on
  the general properties of the diffuse hot baryons in galaxy clusters
  and groups: scaling relations, temperature, entropy and pressure
  radial profiles, and ICM chemical enrichment. 
 We find that simulations including AGN feedback produce scaling
  relations between X-ray observable quantities that are in good
  agreement with observations at all mass scales.  Observed pressure
  profiles are also shown to be quite well reproduced in our radiative
  simulations, especially when AGN feedback is included. However, our
  simulations are not able to account for the observed diversity
  between cool-core and non cool-core clusters, as revealed by
  X-ray observations: unlike for observations, we find that
  temperature and entropy profiles of relaxed and unrelaxed clusters
  are quite similar and resemble more the observed behaviour of non
  cool--core clusters. As for the pattern of metal enrichment, we find
  that an enhanced level of iron abundance is produced by AGN
  feedback with respect to the case of purely SN feedback. As a
  result, while simulations including AGN produce values of iron
  abundance in groups in agreement with observations, they over--enrich
  the ICM in massive clusters. The efficiency of AGN feedback in
  displacing enriched gas from halos into the inter-galactic medium at
  high redshift also creates a widespread enrichment in the outskirts
  of clusters and produces profiles of iron abundance whose slope is
  in better agreement with observations. By analysing the pattern of
  the relative abundances of silicon and iron and the fraction of
  metals in the stellar phase, our results clearly show that different
  sources of energy feedback leave different imprints in the
  enrichment pattern of the hot ICM and stars. Our results
  confirm that including AGN feedback goes in the right direction of
  reconciling simulation predictions and observations for several
  observational ICM properties. Still a number of important
  discrepancies highlight that the model still needs to be improved to
  produce the correct interplay between cooling and feedback in
  central cluster regions. 

  \end{abstract} 
 

\section{Introduction}

The sensitivity reached by X-ray observations with the current
generation of soft X-ray telescopes (\chandra, \xmm\ and \suzaku)
has provided us with a detailed analysis of the thermo- and
chemodynamical properties of the hot intra-cluster medium (ICM) for
statistically representative samples of galaxy clusters.  These
observations have now stablished some of the main ICM properties:
temperature profiles have negative gradients outside core regions out
to the largest radii covered so far by observations \citep[e.g.,
][]{DeGrandi_2002, Vikhlinin_2005, Zhang_2006, Baldi_2007,Pratt_2007,
  Leccardi_2008}; gas entropy in poor clusters and groups is higher
than expected from simple self-similar scaling relations of the ICM
\citep[see][for a recent review on observational scaling relations and
references therein]{Giodini_2013}; radial profiles of the iron
abundance show negative gradients, more pronounced for relaxed
cool-core (CC) clusters, with central values of \Zfe approaching the
solar abundance and with a global enrichment at a level of about
1/3--1/2 $Z_{Fe,\odot}$ \citep[][]{DeGrandi_2004,Vikhlinin_2005,
  dePlaa_2006,Snowden_2008,Leccardi_2008, Werner_2008}; relaxed
clusters also show core regions with very low amounts of gas cooler
than about one third of their virial temperature
\citep[e.g.,][]{Peterson_2001,Boehringer_2002, Sanderson_2006}; etc.

Complementary to X-ray observations, clusters observed in large
Sunyaev-Zel'dovich (SZ) surveys offer an additional channel to analyze
the properties of the hot ICM \citep[see][for a
review]{Carlstrom_2002}.  While the X-ray signal is proportional to
the square of the gas density, the SZ effect \citep{SZ_1972} depends
on the integrated pressure along the line of sight, which decreases
more gently with radius \citep[e.g.,][]{Arnaud_2010}. It is thanks to
the larger dynamic range in gas density accessible and the redshift
independence of the signal that SZ observations are the ideal
complement to X-ray observations.  New generations of millimetre
instruments are now routinely detecting the SZ signal of galaxy
clusters, sometimes out to large radii, thereby opening a new window
on the study of the thermodynamics of the hot baryons in galaxy clusters
\citep[e.g.,][]{Planck_2013, SPT_2013, ACT_2013, Carma_2013}.

Given the range of scales involved by galaxy clusters, their
observational properties arise from a non-trivial interplay between
gravitational processes, which shape the large-scale structure of the
Universe, and a number of astrophysical processes that take place on much
smaller scales (e.g., radiative cooling, star formation and its
associated energy and chemical feedback and AGN heating).  Within this
context, it is only with cosmological hydrodynamical simulations that
one can capture the full complexity of the problem \citep[see][for
recent reviews]{Borgani_2009, Kravtsov_2012}.  In the last two decades
much progress has been made in the numerical modelling of the
formation and evolution of galaxy clusters and groups, thanks to the
ever evolving efficiency of sophisticated cosmological hydrodynamical
simulation codes, that include now advanced descriptions of the
astrophysical processes shaping galaxy formation, and the rapid
increase of accessible supercomputing power
\citep[e.g.,][]{Evrard1990, Navarro1995, Bryan1998, Kravtsov_2000,
  Borgani_2001, Springe_2001_b, Kay_2002}.

These simulations have had varying degrees of success in reproducing
the thermodynamical properties of the hot ICM.  It is well established
that simulations that include only the effects of radiative cooling form
too many stars relative to observational results \citep[see][for
further discussion of this `cooling crisis']{Balogh_2001}.  The
solution to this problem should be provided by a suitable mechanism
combining the action of heating and cooling processes in a
self-regulated way \citep[e.g.][for a review]{Voit_2005}.  
Different forms of energy feedback from supernova (SN)
explosions have been proposed to generate a self-regulated
star formation \citep{springel_hernquist03}.  However, models based on
the action of stellar feedback are not able to regulate star formation
to the low levels observed in the most massive galaxies hosted at the
centre of rich galaxy clusters, the so-called brightest cluster
galaxies (BCGs; e.g. \citealt{Borgani_2004} and references therein).
Within this scenario, it is becoming increasingly clear that AGN
feedback is the most plausible source of heating to counteract gas
cooling within the massive DM halos of groups and clusters. It is only
relatively recently that studies of the effect of AGN feedback in
cosmological simulations of galaxy clusters have been undertaken by a
number of independent groups
\cite[e.g.,][]{sijacki_etal07, Puchwein2008, McCarthy_2010,
  Puchwein2010, fabjan_etal10, short_2010, Battaglia2012, 
  Dubois_2011, Martizzi_2012}.  Due to its
central location and its ability to provide enough energy,
a growing number of authors have argued that gas accretion onto
supermassive black holes (SMBHs) plays a crucial role in regulating
the star formation rates of massive galaxies
\citep[e.g.,][]{Granato_2004, springel_etal2005, Bower_2006, Bower_2008} and
suppressing overcooling in groups and clusters
\citep[e.g.,][]{Churazov_2001}, a picture that is broadly supported by
a large body of observational evidence \citep[e.g.][for a
review]{McNamara_Nulsen_2007}.

Motivated by this idea, \cite{springel_etal2005} developed a novel
scheme to follow the growth of supermassive BHs and the ensuing
feedback from AGN  in cosmological smoothed particle
hydrodynamics (SPH) simulations \citep[see also][for subsequent
modifications of this model]{Sijacki_2006,Booth_2009}.  Using this
original implementation of AGN feedback as a reference,
\cite{Sijacki_2006}, \cite{sijacki_etal07}, \cite{Puchwein2008},
\cite{Bhattacharya_2008} and \cite{fabjan_etal10} went one step
further and modified the original implementation to include both
`quasar' and `radio' feedback modes. The former is characterised by
isotropic thermal heating of the gas when accretion rates are high,
while the latter, characteristic when accretion rates are low,
consists of thermal heating meant to mimic `bubbles' observed in many
nearby clusters, which are thought to be inflated by relativistic jets
coming from the central BH\footnote{Only the models by
  \cite{Sijacki_2006} and \cite{sijacki_etal07} were explicitly
  modified to include the possibility to inflate high-entropy bubbles
  in the ICM whenever accretion onto the central BH enters in a
  quiescent `radio' mode.}.  As for Eulerian adaptive mesh refinement
(AMR) simulations, an alternative way of implementing the AGN energy
injection  is through AGN-driven winds, which shock and heat the
surrounding gas 
(e.g., \citealt{Dubois_2011, Gaspari_2011,
 Martizzi_2012};  see also \citealt{Barai_2013} for an
SPH implementation of kinetic AGN feedback). 

As a result of these analyses, AGN feedback was generally found
to yield reduced stellar mass fractions and star formation rates in a
cosmological context, while reproducing better a number of observable
properties of the ICM \citep[see also][]{McCarthy_2010}.  For example,
\cite{Puchwein2008} found that including energy input from
supermassive BHs enables their simulations to reproduce the
luminosity-temperature and gas mass fraction-temperature relations of
groups and clusters.  \cite{fabjan_etal10} found similarly good
matches to these relations, while also showing that their simulations
yielded temperature profiles in reasonable agreement with observations
of galaxy groups.

Less attention has been paid so far to analyze the interplay between
AGN and chemical enrichment by including a detailed chemodynamical
description of the ICM \citep[e.g.,][]{sijacki_etal07, Moll_2007,
fabjan_etal10, McCarthy_2010}.  Measurements of the enrichment
pattern of the ICM represent a unique means towards a unified
description of the thermodynamical properties of the diffuse gas and
of the past history of star formation within the population of cluster
galaxies \citep[e.g., ][]{Renzini_1993,Borgani_2008}.  In fact, recent
analyses of the ICM metal enrichment from cosmological simulations
including the effect of AGN feedback show that this mechanism can
actually displace a large amount of highly-enriched gas that is
already present inside massive halos at the redshift $z\sim 2$--3, at
which SMBH accretion peaks. The resulting widespread enrichment of the
inter-galactic medium leads to a sensitive change of the amount and
distribution of metals within clusters and groups at low redshift
\citep[e.g.,][]{fabjan_etal10,mccarthy_etal11}.

The aim of this paper is to present a detailed analysis of the
properties of the ICM for an extended set of cosmological
hydrodynamical simulations of galaxy clusters, which have been
performed with the TreePM+SPH {\footnotesize {\sc GADGET-3}} code
\citep{springel05}.  We carried out one set of non-radiative
simulations, and two sets of simulations including
metallicity-dependent radiative cooling, star formation, metal
enrichment and feedback from supernovae, one of which also accounts
for the effect of an efficient model of AGN feedback. The scheme of
AGN feedback implemented in these simulations is a modification of the
original model presented in \cite{fabjan_etal10} with some
improvements related, especially, to the way in which BHs are seeded
and are allowed to grow, with their positioning within their host
galaxy and with the way in which the thermal energy is distributed.
We will show results on the effect that the different feedback
mechanisms implemented in our simulations have on the ICM
thermodynamical properties of our systems and on the corresponding
pattern of chemical enrichment.

The paper is organized as follows: in Section~2, we describe our set
of simulated galaxy clusters, the numerical implementation of the
different physical processes included in our simulations, and the
definitions of the observable quantities computed from the
simulations. In Section~3 we present the results obtained from this
set of simulations on general X-ray properties of the ICM and we
compare them with different observational results.  Section~4
is devoted to the description of the metal enrichment of the ICM in
our simulations.  Finally, we discuss our results and summarise our
main conclusions in Section~5.

\section{The simulated clusters}
\label{sec:simulations}

\subsection{Initial conditions}
Our sample of simulated clusters and groups are obtained from 29
Lagrangian regions, centred around as many massive halos identified
within a large-volume, low-resolution N-body cosmological simulation
\citep[see][for details]{Bonafede2011}.  
The cosmological model assumed is a
flat $\Lambda$CDM one, with $\Omega_{\rm{m}} = 0.24$ for the matter
density parameter, $\Omega_{\rm{b}} = 0.04$ for the contribution of          
baryons, $H_0=$~72~km~s$^{-1}$~Mpc$^{-1}$ for the present-day Hubble
constant, $n_{\rm{s}}=0.96$ for the primordial spectral index and
$\sigma_8 = 0.8$ for the normalisation of the power spectrum.  Within
each Lagrangian region we increased the mass resolution and added the
relevant high-frequency modes of the power spectrum, following the
zoomed initial condition (ZIC) technique \citep{Tormen1997}. Outside
these regions, particles of mass increasing with distance from the
target halo are used, so as to keep a correct description of the large
scale tidal field.  Each high-resolution Lagrangian region is shaped
in such a way that no low-resolution particle contaminates the central
halo at $z = 0$ at least out to 5 virial radii\footnote{The virial
  radius, $R_{vir}$, is defined as the radius encompassing the overdensity of
  virialization, as predicted by the spherical collapse model
  \citep[e.g.,][]{Eke1996}.}. As a result, each
region is sufficiently large to contain more than one interesting halo
with no contaminants within its virial radius.

Initial conditions have been generated by adding a gas component only
in the high-resolution region, by splitting each particle into two,
one representing DM and another representing the gas component, with a
mass ratio chosen to reproduce the cosmic baryon fraction. The mass of
each DM particle is $m_{\rm{DM}} = 8.47\times10^8 \, \msun$ and the
initial mass of each gas particle is $m_{\rm{gas}} = 1.53\times10^8
\, \msun$.

\subsection{The simulation models}

Simulations have been carried out with the TreePM--SPH
{\footnotesize {\sc GADGET-3}} code, a more efficient version of the
previous {\footnotesize {\sc GADGET-2}} code \citep{springel05}. In
the high-resolution region gravitational force is computed by
adopting a Plummer-equivalent softening length of $\epsilon =
5\,\rm{h}^{-1}$ kpc in physical units below $z = 2$, kept
fixed in comoving units at higher redshift \citep[see][for an
  analysis of the effect of softening on radiative simulations of
  galaxy clusters]{Borgani_2006}. As for the hydrodynamic
forces, we assume the minimum value attainable by the SPH smoothing
length of the B-spline kernel to be half of the corresponding value
of the gravitational softening length.
We carried out three different sets of simulations, that we tag as  
{\tt \nr}, {\tt \w} and {\tt \agn}, whose description is provided here
below.

\begin{itemize}

\item {\tt \nr}. Non-radiative hydrodynamical simulations, based on
  the entropy-conserving SPH \citep{Springel_2002}, with computation
  carried out using the B-spline kernel with adaptive smoothing
  constrained to attain a minimum value equal to half of the
  gravitational softening scale. The adopted artificial viscosity
  follows the scheme introduced by \cite{Monaghan_1997}, with the
  inclusion of a viscosity limiter, as described by
  \cite{Balsara_1995} and \cite{Steinmetz_1996}. 

\item {\tt \w}. Hydrodynamical simulations including the effect of
  cooling, star formation and SN feedback.  Radiative cooling rates
  are computed by following the same procedure presented by
  \cite{wiersma_etal09}. We account for the presence of the cosmic
  microwave background (CMB) and of UV/X-ray background radiation
  from quasars and galaxies, as computed by \cite{haardt_madau01}. The
  contributions to cooling from each one of eleven elements (H, He, C,
  N, O, Ne, Mg, Si, S, Ca, Fe) have been pre-computed using the
  publicly available {\footnotesize {\sc CLOUDY}} photo-ionisation
  code \citep{ferland_etal98} for an optically thin gas in
  (photo-)ionisation equilibrium.  Once metals are distributed from
  stars to surrounding gas particles, no process is included in the
  simulations to diffuse metals to neighbour gas particles. As a
  consequence, the metallicity field is quite noisy, since heavily
  enriched gas particles may have neighbour particles which are
  instead characterized by low metallicity. This could in turn induce
  a noisy pattern in the computation of the cooling rates. In order to
  prevent such a spurious noise in the computation of radiative
  losses, we decided to smooth the metallicity field, for the only
  purpose of computing cooling rates, by using the same kernel used
  for the SPH computations \citep[see also][]{Wiersma_2010}.

This set of simulations includes star formation and the effect of
galactic outflows triggered by SN explosions. 
As for the star formation model, gas
particles above a threshold density of 0.1 cm$^{-3}$ and below a temperature
threshold of $2.5 \times 10^5$ K are treated as multiphase, so as to provide a
sub-resolution description of the interstellar medium, according to the model
originally described  by \cite{springel_hernquist03}.
Within each
multiphase gas particle, a cold and a hot-phase coexist in pressure
equilibrium, with the cold phase providing the reservoir of star
formation. The production of heavy elements is described by accounting
for the contributions from SNe-II, SNe-Ia  and low and intermediate mass
stars, as described by \cite{tornatore_etal07}. Stars of different
mass, distributed according to a Chabrier IMF \citep{chabrier03},
release metals over the time-scale determined by the mass-dependent
life-times of \cite{padovani_matteucci93}.  Kinetic feedback is
implemented according to the model by \cite{springel_hernquist03}: a
multi-phase star particle is assigned a probability to be uploaded in
galactic outflows, which is proportional to its star formation rate.
We assume $\rm{v}_{\rm{w}} = 500\vel$ for the outflow velocity, while
assuming a mass--upload rate that is two times the value of the star
formation rate of a given particle.

\item {\tt \agn}. Radiative simulations including the same physical
  processes as in the {\tt \w} case, but also the effect of AGN
  feedback.  Our model for the growth of SMBH and related AGN feedback
  is based on the original implementation presented by
  \cite{springel_etal2005} \citep[see also][]{2005Natur.433..604D} and
  quite similar to the one presented in \cite{fabjan_etal10}.  
  Here below we briefly describe the
  BH feedback model used for our simulations, while we refer to
  the recently submitted paper by 
  \cite{RagoneFigueroa_2013} for a more detailed description.  

\begin{itemize}
\item {\it SMBH seeding and growth}. We represent SMBHs with
  collisionless particles, interacting only via gravitational
  forces. During the simulation, we periodically perform the
  identification of DM halos using an on-the-fly Friends-of-Friends
  (FoF) algorithm. When a DM halo is more massive than a given
  threshold $M_{th}$ and does not already contain a SMBH, 
  we seed it with a new BH with an initial small mass of $M_{seed}=5\times10^6$
  h$^{-1}$ M$_\odot$. We set $M_{th}=2.5\times10^{11}$ h$^{-1}$
  M$_\odot$. The SMBH grows with an accretion rate given by the Bondi
  formula \citep{Bondi1952}, and it is Eddington limited\footnote{As explained in the 
  Appendix of  \cite{RagoneFigueroa_2013}, the {\it $\alpha$-modified}
  Bondi accretion rate employed in our model is characterized by an adimensional
  factor $\alpha= 100$.}. For each BH,
  the corresponding Bondi accretion rate is estimated using the local
  gas density assigned at the BH position by the same SPH spline
  kernel used for the hydrodynamic computations. The BH dynamical mass
  is updated according to the accretion rate, but we do not subtract
  the corresponding mass from the surrounding gaseous component. This
  results in a mass non-conservation of the order of at most a
  fraction of percent of the cluster central galaxy stellar mass, but
  improves the positioning of the SMBH particle and, more important,
  avoids the gas depletion in the BH surroundings, that, at the
  resolution of our simulations, would take place on exceedingly large
  scales.

\item {\it SMBH advection}. As recently also discussed by
  \cite{Wurster13} and \cite{Newton13}, numerical effects can drift
  BHs, originally seeded in DM halos, outside such halos. In order to
  pin BHs at the centre of galaxies, at each time step we reposition
  the BH particles at the position of the neighbour particle, of any
  type (i.e. DM, gas or star), which has the minimum value of the
  gravitational potential. We perform the search of such a particle
  within the gravitational softening of the SMBH. 
  When two BHs are within the gravitational softening and their relative velocity
 is smaller than $0.5$ of the sound velocity of the surrounding gas, we 
 merge them and we place the resulting BH at the position of the most massive one.

 Note that we use the gravitational softening as a searching length,
 rather than the BH smoothing length as in most previous
 implementations, for both the advection and the merging algorithms.
 Given the gravitational nature of the numerical processes responsible
 for drifting the BHs away from galaxies, as well as the physical
 processes leading to BH merging, this length scale is the most
 appropriate to be used.  In addition, at the resolution of our
 simulations, the BH smoothing length is unreasonably large for these
 purposes, and significantly larger than the gravitational softening.
 In particular, if we use the former for searching the minimum
 potential particle, BHs dwelling in satellite halos could often be
 spuriously displaced to the center of a more massive DM halo, and
 immediately would merge with the BH dwelling there. This numerical
 over-merging affects the mass function of BHs, thus spuriously
 increasing by merging the number of high-mass BHs, and
 correspondingly depleting the number of BHs at low and intermediate
 masses.  We verified that numerical BH overmerging is significantly
 reduced by repositioning within the gravitational softening length.

\item {\it Thermal Energy distribution}. In our model, the SMBH growth
  produces an energy determined by a parameter $\epsilon_r$ which
  gives the fraction of accreted mass which is converted in
  energy. Another parameter $\epsilon_f$ defines the fraction of
  extracted energy that is thermally coupled to the surrounding gas.
  In our implementation, both of these parameters, $\epsilon_r$ and
  $\epsilon_f$, are given a value of $0.2$.  Finally, we assume a
  transition from a `quasar' phase to a `radio' mode of the BH
  feedback \citep[see also][]{sijacki_etal07,fabjan_etal10}. This
  happens when the accretion rate becomes smaller than a given limit,
  $\dot{M}_{BH}/\dot{M}_{Edd} = 10^{-2}$. In this case, we increase
  the feedback efficiency $\epsilon_f$ by a factor of four. In the
  original implementation by \cite{springel_etal2005}, the thermal energy
  from the BHs was simply added to the specific internal energy of gas
  particles. As a consequence, whenever this external energy was given
  to a star-forming gas particle, it was almost completely lost. This
  happens because in the effective model of star formation and
  feedback \citep{springel_hernquist03}, the internal energy of
  star-forming particles converges to the equilibrium energy of the
  inter-stellar medium on a very short time scale, and the equilibrium
  energy is independent on the specific internal energy of the gas
  \citep[see discussion in][]{RagoneFigueroa_2013}. To avoid this,
  whenever a star-forming gas particle receives energy from a SMBH, we
  now calculate the temperature at which it would heat the cold gas 
  phase\footnote{The AGN energy is given to the hot and cold phase
    in proportion to their mass.}. If this temperature is larger than
  the average temperature of the gas particle (before receiving AGN
  energy), we consider the particle not to be multi-phase anymore and
  prevent it from forming stars. The particle is, however, subject to
  normal radiative cooling. We also add a temperature threshold of
  $5\times 10^4$ K as a further condition, besides density, for a gas
  particle to become multi-phase and form stars. This is needed
  because in our new scheme a dense gas particle can be very hot: if
  it became star-forming again, it would immediately lose all of its
  internal specific energy in excess to the equilibrium one.
\end{itemize}
  
\end{itemize}

\subsection{Identification of clusters}
The identification of clusters proceeds by running first a FoF
algorithm in the high-resolution regions, which links DM particles
using a linking length of 0.16 times the mean interparticle
separation. The centre of each halo is then identified with the
position of the DM particle, belonging to each FoF group, having the
minimum value of the gravitational potential.  Starting from this
position, and for each considered redshift, a spherical overdensity
algorithm is employed to find the radius $R_{\Delta}$ encompassing a
mean density of $\Delta$ times the critical cosmic density at that
redshift, $\rho_c(z)$.  In the present work, we consider values of the
overdensity\footnote{The corresponding radii approximately relate to
  the virial radius as $(R_{2500}, R_{500}, R_{180}) \approx (0.2,
  0.5, 0.7)\,R_{\rm vir}$ \citep[e.g.,][]{Ettori2006}.}
$\Delta=2500$, $500$ and $180$.  For the sake of completeness, we also
consider the virial radius which defines a sphere enclosing the virial
density $\Delta_{\rm vir}(z)\rho_c(z)$, predicted by the spherical
collapse model: $\Delta_{\rm vir}\approx 93$ at $z=0$ and $\approx
151$ at $z=1$ for the cosmological model assumed by our simulations \citep{Bryan1998}.

\begin{figure}
\centerline{\includegraphics[width=8cm]{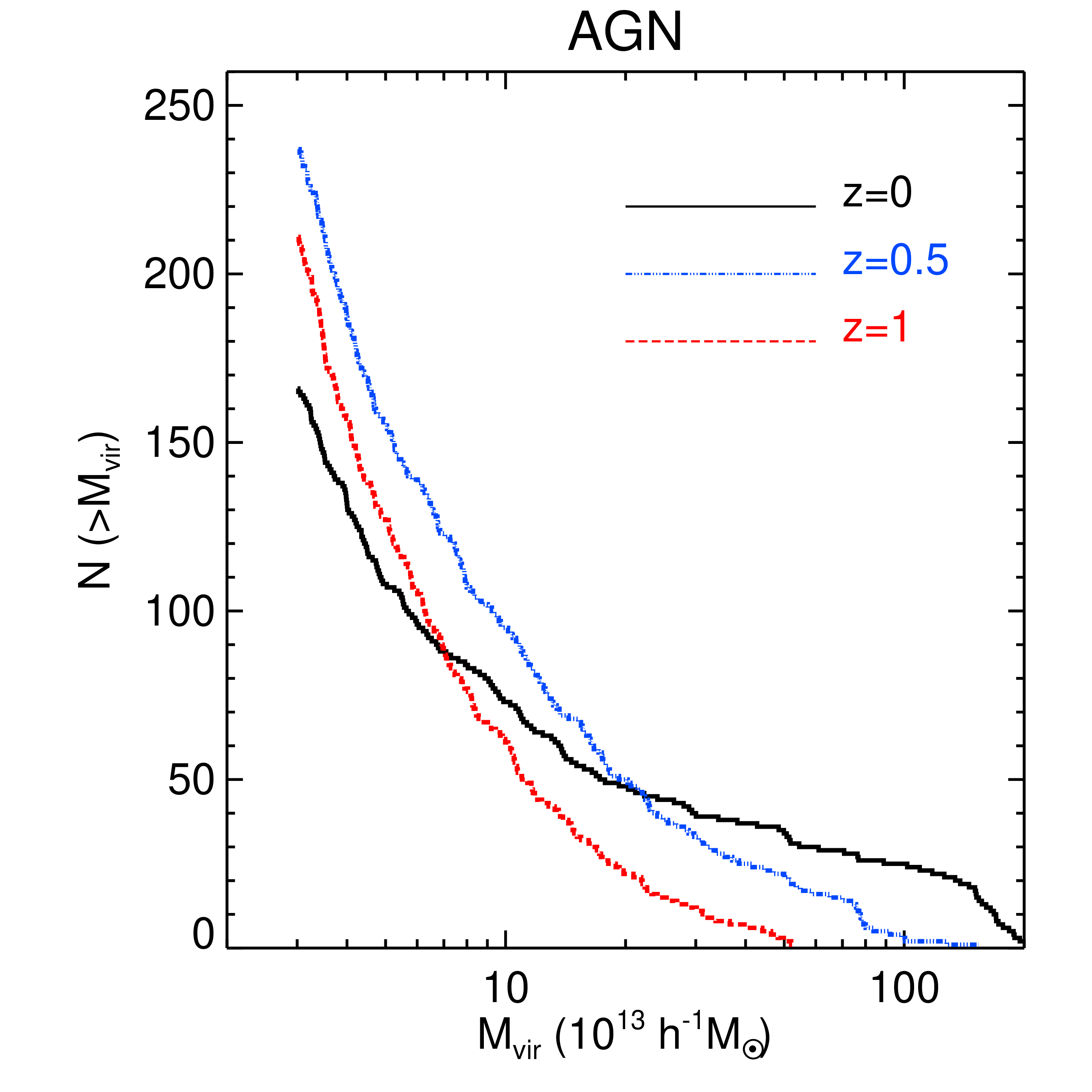}}
\caption{Cumulative distribution of masses $M_{vir}$ for the set of
  clusters and groups identified in the {\tt \agn} set of
  simulations. Solid black, dot-dashed blue and dashed red lines
  correspond to redshifts $z=0, 0.5$ and 1, respectively.}
\label{fig:statistics}
\end{figure}

In total, we end up with a sample of about 160 clusters and groups
having $M_{vir}>3\times 10^{13}h^{-1}M_\odot$ at $z=0$. 
This number is larger at higher redshift: $\sim 240$ systems at $z=0.5$ and
$\sim 200$ at $z=1$.
In Fig.~\ref{fig:statistics} we show the cumulative number of clusters 
within the {\tt \agn} set of simulations as
a function of their mass $M_{vir}$  at $z=0, 0.5, 1$.

\subsection{Computing the observable quantities}

In order to analyse the results of our simulations, we will study
several thermodynamical and chemodynamical properties which can be
directly compared with observational data, and which have been widely
studied by different simulations.  Here we briefly
describe how these quantities have been computed from our simulations.

\begin{itemize}
\item  X-ray luminosity. This quantity is computed by summing the
contributions to the emissivity, $\epsilon_i$, carried by all the gas
particles within the region ${R_\Delta}$ where $L_X$ is computed: 
\be
L_X=\sum_i\epsilon_i=\sum_i\,n_{{\rm e},i}n_{{\rm H},i}\Lambda(T_i,Z_i){\rm d}V_i\,,
\label{eq:emiss}
\ee
where $n_{{\rm e},i}$ and $n_{{\rm H},i}$ are the number densities of electrons
and of hydrogen atoms, respectively, associated with the $i$-th gas
element of given density $\rho_i$, temperature $T_i$, mass $m_i$,  metallicity
$Z_i$, and  volume ${\rm d}V_i=m_i/\rho_i$. 
Furthermore, $\Lambda(T,Z)$ is the temperature- and
metallicity-dependent emissivity computed within a given energy band.
 The energy--dependent emissivity for each particle is calculated
  using the plasma emission model by \cite{RaymondSmith77}. The
  spectra are computed by binning the emissivity within energy
  intervals, so as to have an energy resolution $\Delta E =
  0.01$. Luminosities within a given energy band are then computed by
  integrating this energy--dependent emissivity within the appropriate
  energy interval. We point out that the assumed cooling function
  adopted for the computation of X--ray luminosity depends on a global
  metallicity, rather than accounting for the contributions from all
  the chemical species followed in our simulations. In the
  following, we will present results on the X-ray luminosity as computed
  within the $[0.1-2.4]\,$keV energy band.

\item Temperature.
Different proxies to the X-ray observational
definition of temperature have been proposed in the literature.  
In general, the ICM temperature can be estimated as
\be
T\,=\,{\sum\limits_i^{} w_i T_i\over \sum\limits_i^{} w_i}\,,
\label{eq:tgen}
\ee
where $T_i$ is the temperature of the $i$-gas element, which
contributes with the weight $w_i$.  The mass-weighted definition of
temperature, $T_{\rm mw}$, is recovered for $w_i=m_i$, which also
coincides with the electron temperature for a fully ionised
plasma. The emission-weighted temperature would be instead recovered
for $w_i=\epsilon_i$.
On the contrary, the spectroscopic-like estimate of temperature,
$T_{\rm sl}$, is recovered from Eq.~\ref{eq:tgen} by using the weight
$w_i=\rho_i m_i T^{\alpha-3/2}$ with $\alpha=0.75$
\citep{Mazzotta_2004}.  In the Bremsstrahlung regime ($T\gtrsim2-3$
keV), this temperature estimator has been shown to provide a close
match to the actual spectroscopic temperature, obtained by fitting
X-ray spectra of simulated clusters with a single-temperature plasma
model.  When measuring the spectroscopic-like temperature \tsl\ of our
systems, we follow the procedure described by \cite{Vikhlinin_2006_b},
which generalizes the analytic formula originally introduced by
\cite{Mazzotta_2004} to include relatively cold clusters with
temperature below 3 keV.

\item  Entropy. This is another useful quantity to characterise the thermodynamical status
of the ICM \citep{Voit_2005}. 
We use the standard definition of entropy
usually adopted in X-ray studies of galaxy clusters:
\begin{equation}
K_\Delta\,=\,k_B T_\Delta n_{e,\Delta}^{-2/3}\,, 
\label{eq:entr}
\end{equation} 
where $n_{e,\Delta}$ is the electron number density computed at
$R_\Delta$, and $k_B$ is the Boltzmann constant.

\item Total thermal content. A useful proxy to the total thermal
  content of the ICM is the quantity $Y_X=M_gT$, with $M_g$ being
  the gas mass and $T$ one estimator of the global ICM temperature
  \citep[e.g.,][]{Kravtsov_2006}.  In our case, we will use two
  slightly different definitions of $Y_X$, the mass-weighted, $Y_{X,
    mw} =M_gT_{mw}$, and the spectroscopic-like, $Y_{X, sl}
  =M_gT_{sl}$, estimates. Being a proxy of the total thermal content
  of the ICM, $Y_X$ turns out to be also a robust mass proxy, whose
  scaling relation against total cluster mass has a low scatter and is
  almost independent on the physical processes included in the
  simulations \citep{Kravtsov_2006, Stanek_2010, Fabjan_2011}. In order
  to minimize the contribution to the scatter from the cluster central
  regions, \cite{Kravtsov_2006} showed that the temperature should be
  estimated by excising the regions within $0.15R_{500}$. We will also
  adopt this procedure in the following. 

\item  Pressure. By assuming an ideal gas equation of state, we
  compute the volume-weighted estimate of
the gas pressure as
\be
P\,=\,{\sum\limits_i^{} p_i dV_i\over \sum\limits_i^{} dV_i}\,,
\label{eq:pgen}
\ee
where $p_i=(k_B/ \mu m_p)\rho_i T_i$ and $dV_i$ are the contributions
to pressure and volume, respectively, of each considered gas particle
($\mu$ and $m_p$ being the mean atomic weight and the proton mass,
respectively).

\item  Metallicity of the ICM. From an observational point of view,
  this quantity is computed
through a spectral fitting procedure, by measuring the equivalent
width of emission lines associated with a transition between two
heavily ionised states of a given element. The simplest
proxy to this spectroscopic measure of the ICM metallicity is,
therefore, the emission-weighted definition,
\be 
Z_{\mathrm{ew}}\,=\,{\sum_i Z_i m_i \rho_{i}\Lambda(T_i,Z_i)\over
  \sum_i m_i \rho_{i}\Lambda(T_i)}\,,
\label{eq:z_ew}  
\ee
where $Z_i$, $m_i$, $\rho_{i}$ and $T_i$ are the
metallicity, mass, density and temperature of the $i$--gas element,
with the sum being performed over all the gas particles lying within
the cluster extraction region. \cite{Rasia_2008} showed that the
emission-weighted estimator of Eq.~\ref{eq:z_ew} provides values of
metallicity that are quite close to the actual spectroscopic ones, at
least for iron and silicon, while abundance of oxygen can be severely
biased in high-temperature, $T\magcir 3$ keV, systems. 

Since both simulated and observed
metallicity radial profiles are characterised by significant negative
gradients, we expect the ``true'' mass-weighted metallicity,
\be 
Z_{\mathrm{mw}}\,=\,{\sum_i Z_i m_i\over \sum_i m_i}\,,
\label{eq:z_mw}  
\ee
to be lower than the ``observed'' emission-weighted estimate.

\end{itemize}

\section{General X-ray properties}
\label{sec:xproperties}

Before presenting the general X-ray properties of our sample of galaxy clusters, we highlight 
results from previous works that analyze different aspects of the same suite of simulations. 
Together, these results establish a reasonable level of consistency with observations with 
regards to the baryonic content of the clusters, emphasize the important role of radiative and 
feedback processes, and provide important calibrations for observational tests.

Generally speaking, as shown in  \citet{Planelles_2012}, including the effect of AGN feedback 
helps to alleviate tension with observations for the stellar, the hot gas and the total baryon mass 
fractions. However, both of our radiative simulation sets predict a trend for the stellar mass fraction 
with cluster mass that is weaker than observed. On the other hand, this tension depends on the 
particular set of observational data considered.
For massive clusters, the ratio between the cluster baryon content and the cosmic baryon fraction, 
$Y_b$, when computed at $R_{500}$, is nearly independent of the physical processes included and 
characterized by a negligible redshift evolution. At smaller radii, i.e. $R_{2500}$, the typical value of 
$Y_b$ slightly decreases, by an amount that depends on the physics included in the simulations, while 
its scatter increases by a factor of two.  These results have interesting implications for the cosmological
applications of the baryon fraction in clusters.
  
Some of the tensions with observations may be due to difficulties in accounting for the diffuse stellar content, 
distinct from cluster member galaxies.
\citet{Cui_2013} have carried out a detailed analysis of the performance of two different methods to identify 
the diffuse stellar light. One of the methods separates the BCG from the `diffuse stellar component' (DSC) 
via a dynamical analysis. The other is inspired by the standard observational technique: mock images are 
generated from simulations, and a standard surface brightness limit (SBL) is assumed to disentangle the BCG 
from the intra-cluster light (ICL). Significant differences are found between the ICL and DSC fractions computed 
with these two methods. The use of a brighter SBL can reconcile the ICL and DSC fractions but the exact value, 
as calibrated by the simulations, is quite sensitive to feedback.
  
 Moreover, \citet{RagoneFigueroa_2013} investigated some problems that are common throughout the literature 
 of numerical simulations of galaxy clusters that also include the effect of AGN feedback. For example, the inclusion 
 of AGN feedback helps to reduce the stellar mass content of BCG galaxies, yet their stellar masses remain three 
 times higher than observed. Correspondingly, there is still some tension between predictions and observations of 
 the structural properties of BCGs, namely, larger half-light radii and indications of a flattening of the stellar density 
 profiles on scales \gtrsim $10\,kpc$, much larger than observed.  In addition, the BCGs stellar velocity dispersions 
 are too large, but the implementation of the AGN feedback has little impact here.

\subsection{Self-similar scaling relations}

The simplest model to describe the properties of the ICM is the
self-similar model originally derived by \cite{Kaiser_1986}. This
model assumes an Einstein--de Sitter background cosmology, that the
shape of the power spectrum of fluctuations is strictly a power law and
that no characteristic scales are introduced in the collapse process
that leads to cluster formation. The first two assumptions amount to
assume that no characteristic scale is present in the underlying
cosmological model. The third assumption is naturally satisfied in the
case that gravity is the only process driving halo collapse and gas
heating. The self-similar model further assumes clusters to be
spherically symmetric and in hydrostatic equilibrium (HE). A further
key assumption of this model is that the logarithmic slopes of gas
density and temperature profiles, which enter in the HE equation to
provide the mass within a given radius, are independent of the cluster
mass \citep[see][for a recent review]{Kravtsov_2012}.  In such
self-similar model several simple relations between different X-ray
properties of the gas in clusters can be predicted.  Here we provide
some of the scaling relations that will be relevant for our analysis.

If, at redshift $z$, $M_{\Delta}$ is the mass contained within the
radius $R_{\Delta}$, enclosing a mean overdensity $\Delta$ times the
critical cosmic
density\footnote{M$_\mathrm{\Delta}=\Delta\,\rho_{c}(z)\,(4\,\pi/3)R_{\Delta}^3$,
  where $\rho_{c}(z)=3H_0^2E(z)^2/8\pi G$ is the critical density and
  $E(z)\,\equiv \,H(z)/H_0\,=\,\left[(1+z)^3\Omega_m+\Omega_\Lambda\right]^{1/2}$
  in a spatially-flat $\Lambda$CDM cosmological model.}, the total
enclosed mass scales with temperature as:
\begin{equation}
M_{\Delta}\,\propto \,T_\Delta^{3/2}E^{-1}(z)\, ,
\label{eq:mt_ss}
\end{equation}
where $T_\Delta$ is the gas temperature measured at $R_\Delta$.  This
relation between mass and temperature can be turned into scaling
relations among other observable quantities.

Assuming that the gas distribution traces the dark matter distribution
and that the thermal Bremsstrahlung process dominates the emission
from the ICM plasma, the X-ray luminosity scales as
 \begin{equation}
L_{X,\Delta}\propto M_{\Delta}\rho_c T_\Delta^{1/2} \propto T_\Delta^2E(z) .
\label{eq:lx_t} 
\end{equation}
Within the self-similar model, the entropy computed at a fixed
overdensity $\Delta$ (see Eq. \ref{eq:entr}), scales with temperature
and redshift according to
\begin{equation}
K_{\Delta}\propto T_\Delta E^{-4/3}(z)\,.
\label{eq:entr_ss}
\end{equation}
As for $Y_X$, its self-similar scaling against mass computed within
$R_\Delta$ is predicted to be
\begin{equation}
Y_{X,\Delta}\propto M_\Delta^{5/3} E^{2/3}(z)\,.
\label{eq:yx_m}
\end{equation}

A number of observations of representative samples of galaxy clusters
in the local Universe have established that scaling relations
predicted by the self-similar model do not match the observational
results. For instance, the relation between X-ray luminosity and mass
is steeper than the self-similar prediction
\citep[e.g.,][]{Chen_2007}.  Consistently with the $L_X$-$M$ relation,
the observed $L_X$-$T$ scaling is also steeper than predicted
\citep[e.g.,][]{Markevitch_1998, Arnaud_1999,Osmond_2004, Pratt_2009},
$L_X\propto T^{\alpha}$ with $\alpha \simeq 2.5-3$ for clusters
($T\gtrsim 2$ keV) and possibly even steeper for groups ($T\mincir 1$
keV).  Furthermore, the measured gas entropy in central regions is
higher than expected \citep[e.g.,][]{Sun2009,Pratt_2010}, especially
for poor clusters and groups, with respect to the $K\propto T$
predicted scaling.  This extra entropy prevents the gas from being
compressed to high densities, reducing its X-ray emissivity compared
to the self-similar prediction.

These discrepancies between the self-similar model and the
observations have motivated the idea that, besides gravity, some
important physics related with the baryonic component is missing in
the model.  The main non-gravitational physical processes thought to
be responsible for boosting the entropy of the ICM are heating from
astrophysical sources, such as SNe and AGN, and the removal of
low-entropy gas via radiative cooling
\citep[e.g.][]{Voit_2005}. Lower-mass systems are affected more than
massive objects, thus breaking the self-similarity of the scaling laws
and providing, therefore, a probe of the star formation and feedback
processes operating in cluster galaxies.

\subsection{Scaling relations at $z=0$}

In this Section we analyse the scaling relations for our sample of
simulated clusters and groups, paying special attention to the effect
that the different physical models have on them. In particular, we analyse the
$L_X-T$, 
$K-T$, 
$Y_X-M$ and $T-M$ relations at $z=0$.  Only clusters with
$M_{vir}\gtrsim 3\times 10^{13}\msun$ within our {\tt \nr}, {\tt \w},
and {\tt \agn} simulation sets are included in our analysis.  Unless
otherwise stated, we compute these scaling relations adopting
$\Delta=500$ and, therefore, all quantities are evaluated within
$R_{500}$.  The reason for this choice is that, typically, $R_{500}$
corresponds to the most external radius out to which detailed X-ray
observations \citep[e.g.,][]{Maughan_2007, Vikhlinin_2009, Sun2009,
  Pratt_2009}, possibly in combination with SZ observations
\citep[e.g.][]{Planck2011}, are provided.

We will use the \tsl\ estimator for the temperature in all the
comparisons with X-ray data.  However, it is well known that,
compared with the mass-weighted temperature, the \tsl\ produces larger
scatter in the scaling relations \citep[e.g.,][]{Fabjan_2011}, owing to
its sensitivity to the contribution of the cold component of the gas
temperature distribution.  In order to analyze the intrinsic scatter
in these relations and their behaviour as mass proxies, we will
compute them by using both estimates of the ICM temperature, that is,
$T_{sl}$ and $T_{mw}$.  Correspondingly, we will show the $Y_X-M$
relation for both estimates of $Y_X$ ($Y_{X, sl}$ and $Y_{X, mw}$).
In addition, in order to reduce the scatter in the scaling relations
involving temperature \citep[e.g.,][and references
therein]{Pratt_2009} and to better reproduce the procedure of
observational analyses, we exclude the central regions,
$r<0.15R_{500}$, in the computation of the temperature.

For each set of cluster properties, 
$(X,F)=(T, L_X)$, $(T, K)$,  $(M, Y_X)$ and $(M, T)$, 
we fit to our sample of simulated clusters at $z=0$ a 
power-law scaling relation of the form 
\begin{equation} 
\label{eq:genscalerel}
F=C_0\left(\frac{X}{X_0}\right)^{\alpha},
\end{equation}
by minimising 
the unweighted $\chi^2$ in log space. Here $X_0=6\ {\rm keV}$ if $X=T$
and $X_0=5.0\times 10^{14}\msun$ if $X=M$.  The best-fitting
parameters $\alpha$ and $C_0$ for each relation are summarised in
Table \ref{tab:scalerelz0}.  Following a similar approach to that
presented in \citet{short_2010}, the scatter in these relations,
$\sigma_{\log_{10}F}$, is estimated as the rms deviation of
$\log_{10}{F}$ from the mean relation:
\begin{equation}
\label{eq:sig}
\sigma_{\log_{10}{F}}^2=\frac{1}{N-2}\sum_{i=1}^{N}\left[\log_{10}{F_i}-\alpha\log_{10}{\left(\frac{X}{X_0}\right)-\log_{10}{C_0}}\right]^2,
\end{equation} 
where $N$ is the number of individual data points. 
The scatter about each relation is also listed in Table \ref{tab:scalerelz0}.

\begin{table}
  \caption{Best-fit parameters (with $1\sigma$ errors) for the X-ray scaling relations for our 
    sample of clusters in the {\nr}, {\w} and {\agn} sets at
    $z=0$. Only clusters with $M_{vir}\gtrsim 3\times 10^{13}\msun$
    have been considered. All quantities have been computed within $R_{500}$.
    For each set of cluster properties, a power-law scaling relation given by Eq. \ref{eq:genscalerel} 
    is fit to our sample of  simulated clusters.  According to this fitting, $C_0$, $\alpha$ and 
    $\sigma_{\log_{10}{F}}$ (see Eq. \ref{eq:sig}) represent the normalisation, 
    the slope and the scatter of the different relations, respectively.
    The normalisation $C_0$ has units of $10^{44} {\rm erg\ s}^{-1}$, $keV cm^2$, $10^{14} \msun keV$,
    and $keV$ for  $L_X-T$, $K-T$, $Y_X-M$, and $T-M$, respectively.
    }
 \begin{tabular}{@{}lccc}
\hline
Relation & $C_0$ & $\alpha$ & $\sigma_{\log_{10}F}$ \\
\hline
{\tt \nr} simulation & & & \\
\tslm & $3.24\pm 0.07$ & $0.51\pm 0.01$ & $0.07$\\
\tmwm & $4.89 \pm 0.06$ & $0.65 \pm 0.01$  & 0.03 \\
\yxmsl & $2.34\pm 0.06 $ & $1.53\pm 0.01$ & $0.09$\\
\yxmwm & $3.53 \pm 0.06$ & $1.67 \pm 0.01$  & 0.05 \\
\lxtslband & $61.08\pm 4.53$ & $2.29\pm 0.04$ & $0.16$\\
\lxtmwband & $23.16 \pm 1.25$  & $1.85 \pm 0.03$  & $0.15$ \\
\stsl & $2588.30\pm 99.21$ & $1.17\pm 0.02 $ & $0.08$\\
\stmw & $1550.44 \pm 52.64 $ & $0.93 \pm 0.02$  & 0.09 \\

\\
{\tt \w} simulation & & & \\
\tslm & $4.80\pm 0.08$ & $0.55\pm 0.01$ & $0.05$\\
\tmwm & $5.40 \pm 0.08$ & $0.60 \pm 0.01$  & 0.05 \\
\yxmsl & $2.55\pm 0.06$ & $1.66\pm 0.01$ & $0.08$\\
\yxmwm & $2.87 \pm 0.06$ & $1.71 \pm 0.01$ & 0.07 \\
\lxtslband & $7.49 \pm 0.61$ & $2.17\pm 0.05$ & $0.22$\\
\lxtmwband & $5.80 \pm 0.43 $ & $2.00  \pm 0.05$ & $0.21$ \\
\stsl & $2027.88\pm 42.16$ & $1.02\pm 0.01$ & $0.06$\\
\stmw & $1787.95 \pm 37.55$ & $0.94 \pm 0.01$  & 0.06 \\
\\
{\tt \agn} simulation & & & \\
\tslm & $5.02\pm 0.07$ & $0.54\pm 0.01$ & $0.04$\\
\tmwm & $5.23 \pm 0.07$ & $0.55 \pm 0.01$  & 0.04 \\
\yxmsl & $3.19\pm 0.08$ & $1.73\pm 0.01$ & $0.08$\\
\yxmwm & $3.32 \pm 0.08$ & $1.74 \pm 0.01$  & 0.07 \\
\lxtslband & $9.48\pm 0.77$ & $2.46\pm 0.05$ & $0.23$\\
\lxtmwband &  $8.56 \pm 0.67$ & $2.43  \pm 0.05$  & $0.22$  \\
\stsl & $1686.51 \pm 33.27$ & $0.94\pm 0.01$ & $0.05$\\
\stmw & $1619.83 \pm 31.76$ & $0.93 \pm 0.01$  & 0.06 \\
\hline
\end{tabular}

\label{tab:scalerelz0}
\end{table}

In order to compare with observational data, we use a selection of
representative observational samples, mainly from \citet{Pratt_2009},
\citet{Pratt_2010}, \citet{Sun2009}, 
\citet{Mahdavi_2013} and \citet{Eckmiller_2011}.  To `scale' these
observational data to $z=0$ for comparison with our simulated
clusters, the correction factor $E(z)^n$ is included, thus removing the
self-similar evolution predicted by
Eqs. (\ref{eq:lx_t})--(\ref{eq:yx_m}).

Before presenting the results of our analysis, we briefly describe the
main characteristics of each of the observational data sets we compare
with.
\cite{Pratt_2009} and \cite{Pratt_2010} examine the X-ray properties
of 31 nearby galaxy clusters from the Representative \xmm\ Cluster
Structure Survey \citep[REXCESS,][]{Bohringer_2007}. This sample,
which includes clusters with temperatures in the range 2--9 keV, has
been selected in X-ray luminosity only, with no bias towards any
particular morphological type. According to their central densities,
clusters in this sample have been classified as relaxed, cool-core
(CC) systems or as morphologically disturbed or non-cool core (NCC)
systems.  These data are particularly suitable for a comparison with
our simulated cluster samples because spectral
temperatures, luminosities, masses and entropies are tabulated within $R_{500}$.

\citet{Sun2009} present an analysis of 43 nearby galaxy groups
($kT_{500} = 0.7 - 2.7$ keV or $M_{500} = 10^{13} - 10^{14} h^{-1}$
M$_{\odot}$, 0.012 $<z<$ 0.12), based on \chandra\ archival data.
They trace gas properties out to at least $R_{2500}$ for all 43
groups. For 11 groups, gas properties are robustly derived to
$R_{500}$ and, for an additional 12 groups, they derive properties at
$R_{500}$ from extrapolation.

In a recent work within the Canadian Cluster Comparison Project
(CCCP), \citet{Mahdavi_2013} present a study of multiwavelength X-ray
and weak lensing scaling relations for a sample of 50 clusters of
galaxies in the redshift range $0.15 < z < 0.55$.  After considering a
number of scaling relations, they found that gas mass is the most
robust estimator of weak lensing mass, yielding $15 \pm 6$ per cent
intrinsic scatter at $r_{500}^{WL}$, whereas the pseudo-pressure $Y_X$
yields a consistent scatter of $22 \pm 5$ per cent.

In order to test local scaling relations for the low-mass range,
\citet{Eckmiller_2011} compiled a statistically complete sample of 112 galaxy 
groups from the X-ray selected \emph{HIFLUGCS},
\emph{NORAS}, and \emph{REFLEX} catalogues
\citep[][respectively]{Reiprich_2002, Bohringer_2000, Bohringer_2004}.
Groups were selected by applying an upper limit to the X-ray
luminosity, which was determined homogeneously for all three parent
catalogues, plus a lower redshift cut to exclude objects that were too
close to be observed out to sufficiently large radii.  In this work,
only a subsample of 26 local groups (median redshift $0.025$),
observed with the \chandra\ telescope with sufficient exposure time
($\gtrsim 10\,{ks}$), was investigated.  Temperature, metallicity, and
surface brightness profiles were created for these 26 groups, and used
to determine the main physical quantities and scaling relations.

\subsubsection{Mass scaling relations}

\begin{figure*}
{\includegraphics[width=8cm]{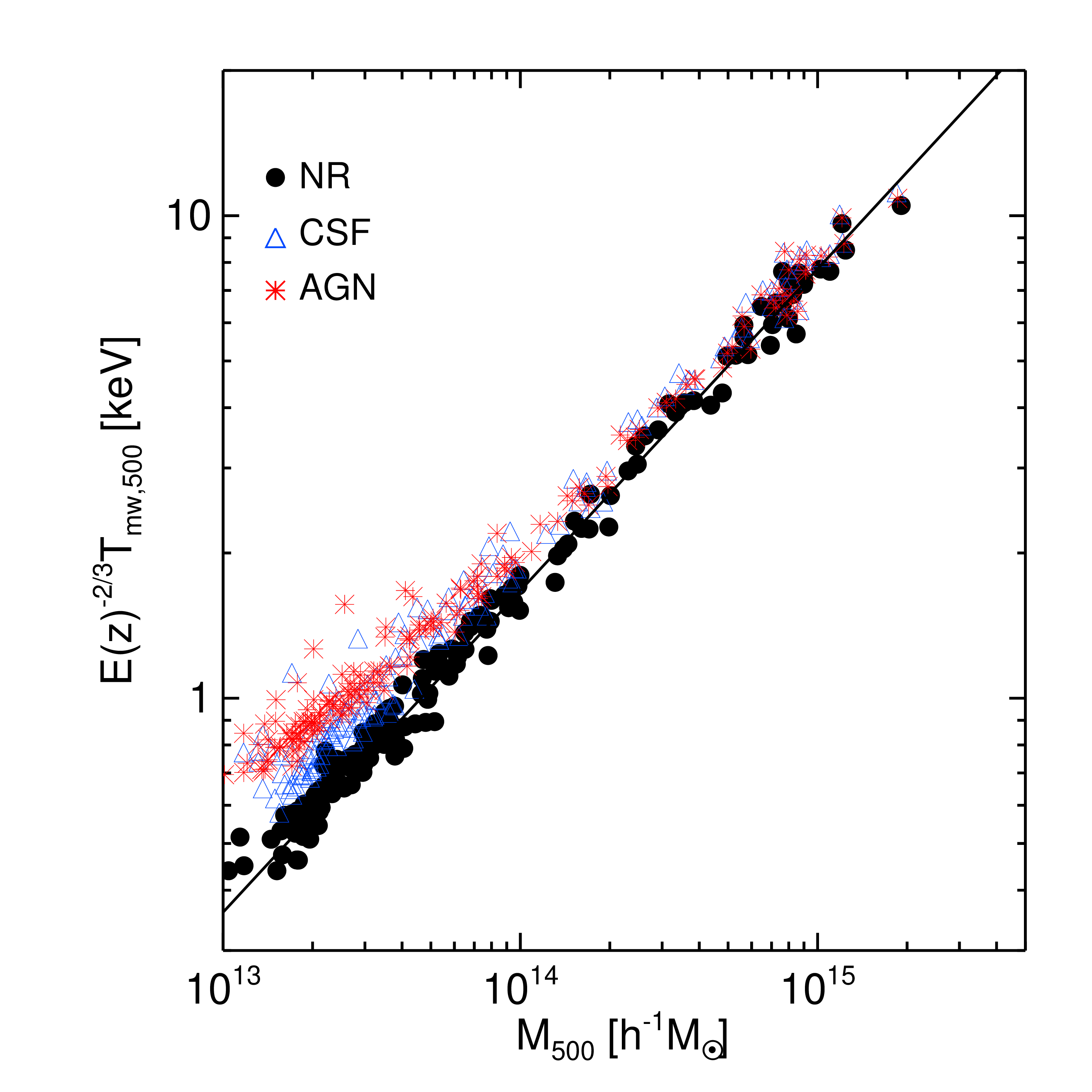}}
{\includegraphics[width=8cm]{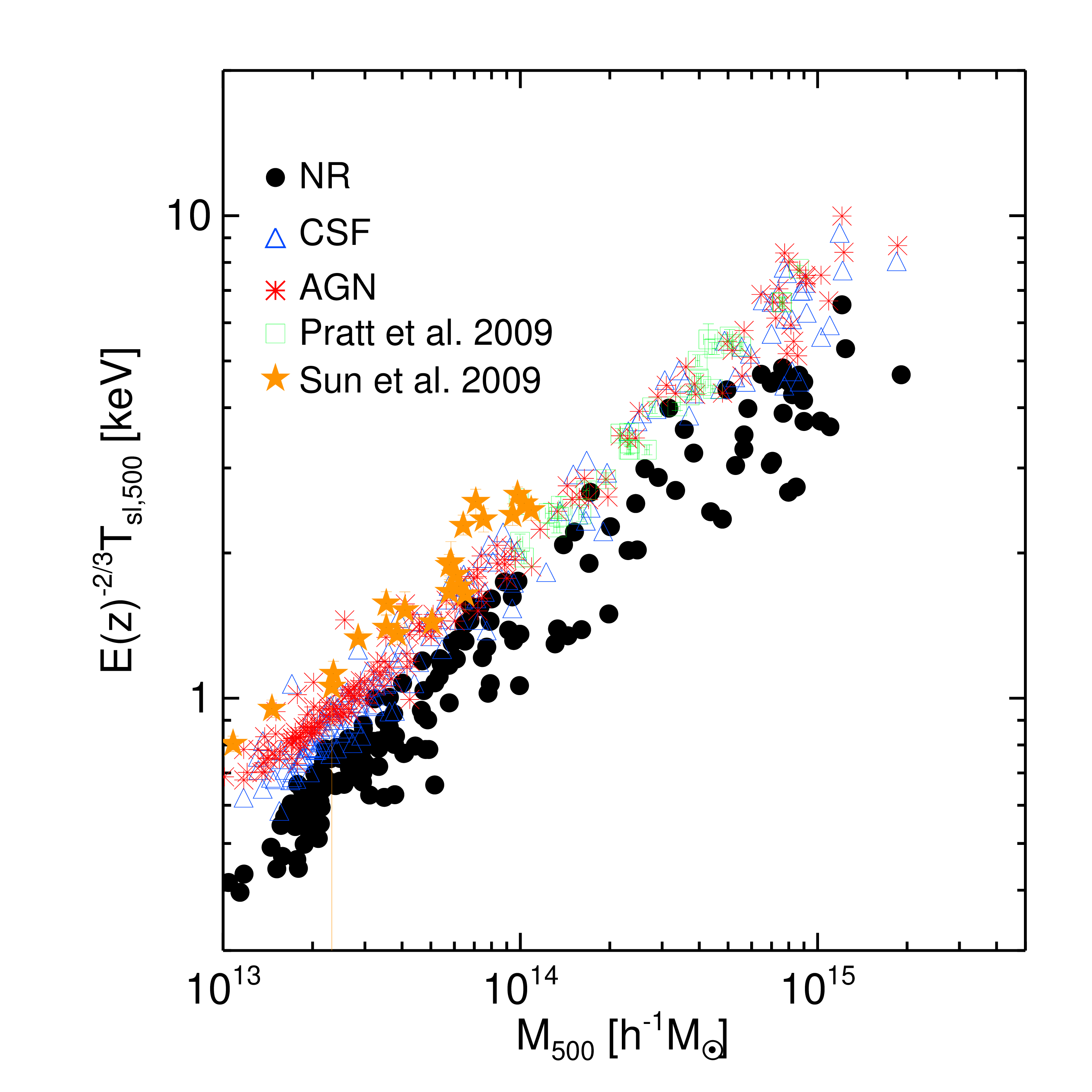}}
\caption{Mass-weighted (left panel) and spectroscopic-like (right
  panel) temperature as a function of total mass within $R_{500}$ at
  $z=0$.  Results for our sample of clusters within the {\tt \nr},
  {\tt \w}, and {\tt \agn} runs as represented by black circles, blue
  triangles and red stars, respectively.  On the left panel, our
  results are compared with the self-similar scaling (black continuous
  line). On the right panel, observational samples from
  \citet{Pratt_2009} and \citet{Sun2009} are used for comparison. }
\label{fig:t_m}
\end{figure*}

Figure \ref{fig:t_m} shows the \tmwm\ and \tslm\ relations (left and
right panels, respectively) within $R_{500}$ for the sample of
clusters in our {\tt \nr}, {\tt \w}, and {\tt \agn} runs. Results
on the \tmwm\ relation are compared with the self-similar prediction,
while results on the \tslm\ relation are compared with observational data from
\citet{Pratt_2009} and \citet{Sun2009}. 

In agreement with previous analyses
\citep[e.g.,][]{Stanek_2010,Fabjan_2011}, the \tmwm\ relation in the
{\tt \nr} case has a slope in close agreement with the self-similar
scaling, $\alpha \sim 0.65\pm 0.01$, also with a tight scatter (see
Table 1).  In addition, whereas at the scale of high-mass systems this
relation is relatively insensitive to baryon physics, at the scale of
groups, $M_{500}\mincir 10^{14}\msun$, a significant deviation from
the self-similar expectation is present in the radiative
simulations. The deviation is more significant for the {\tt \agn}
simulations due to the heating from the BH energy feedback, which is
relatively more efficient in low-mass groups.

As for the comparison with observational data, we consider the
mass--temperature relation using the spectroscopic-like estimator of
temperature. As a word of caution in performing this comparison, we
remind  that we use here true cluster masses for simulations, while
masses for the observational data shown in the right panel of
Fig. \ref{fig:t_m} are based on X-ray data and the application of
hydrostatic equilibrium. While it is beyond the scope of this paper to
carry out a detailed analysis of biases in X-ray mass estimates, the 
evidence from simulation analyses indicates that X-ray masses may 
be underestimated by $\sim 20$ per cent 
\citep[e.g.,][and references therein]{Nagai2007,Rasia_2012}.

We note that in the \tslm\ relation, for a given mass, clusters in the
{\tt \nr} runs are much cooler than observed, by an amount that is
definitely larger than in the \tmwm\ relation.  In addition, the
scatter around the mean relation is larger than in the radiative runs.
By definition (see Section 2.4) the spectroscopic-like
temperature tends to give more weight to the relatively colder component of the
gas temperature distribution \citep[see also][]{Mazzotta_2004}. 
\begin{figure*}
\hbox{
{\includegraphics[width=5.9cm]{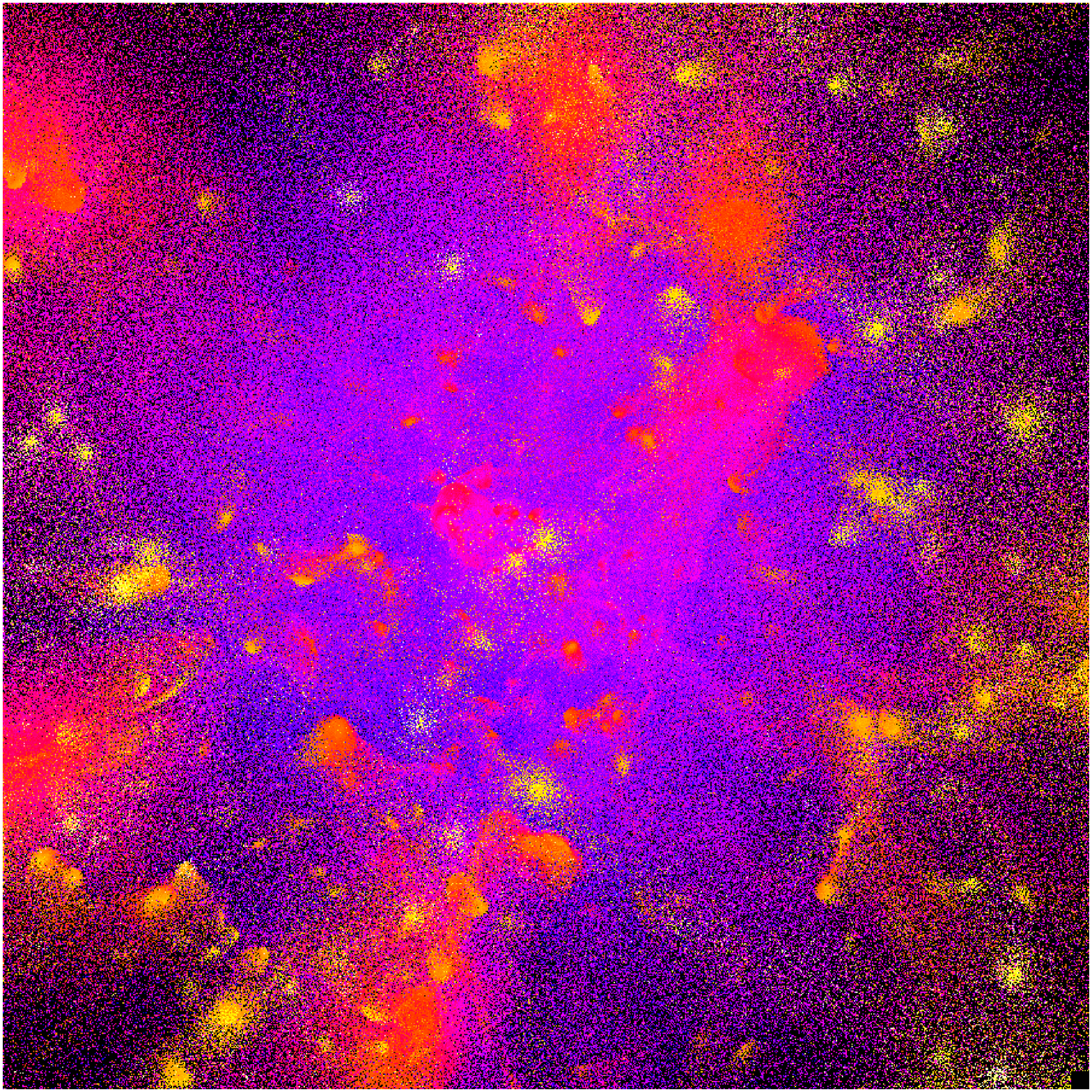}}
{\includegraphics[width=5.9cm]{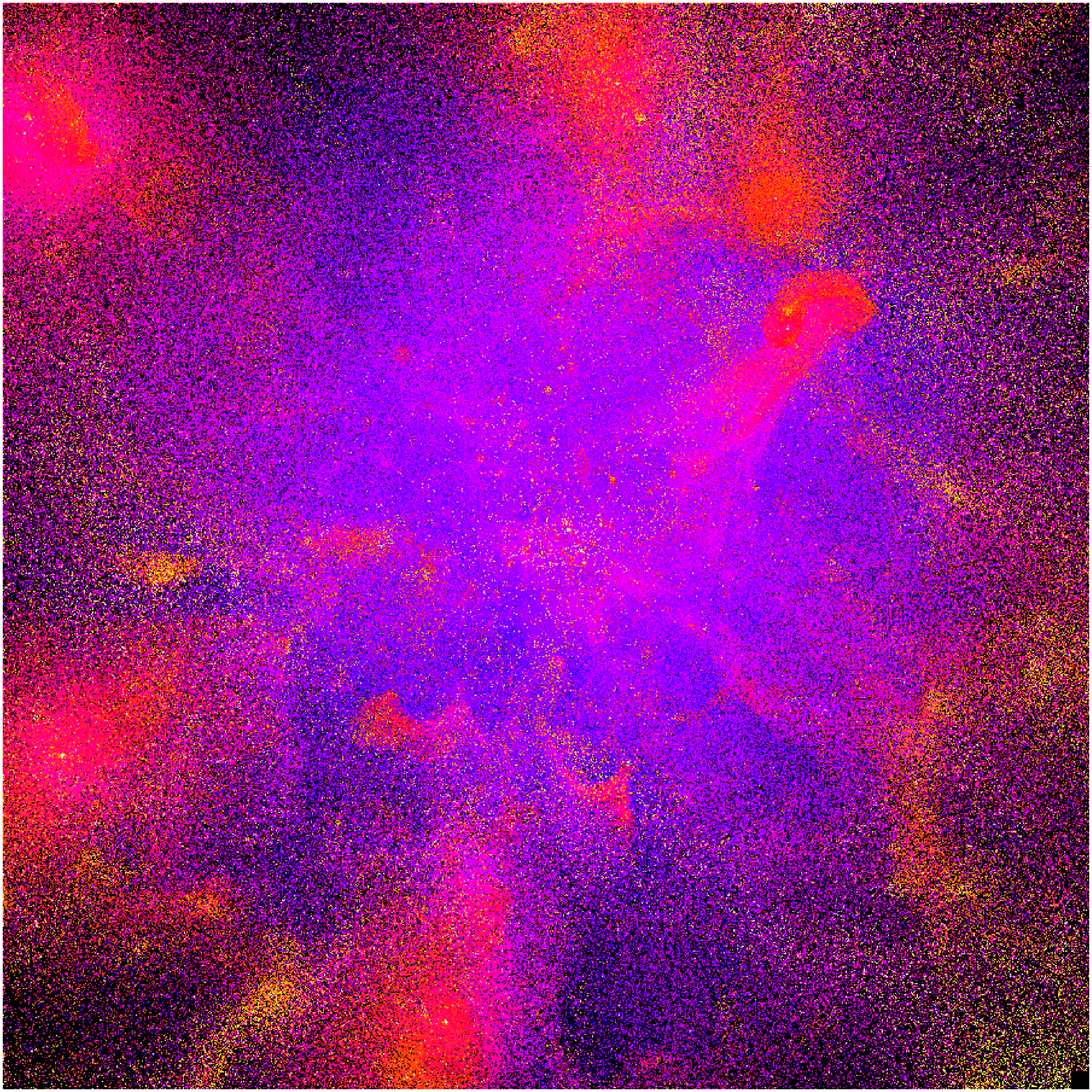}}
{\includegraphics[width=5.9cm]{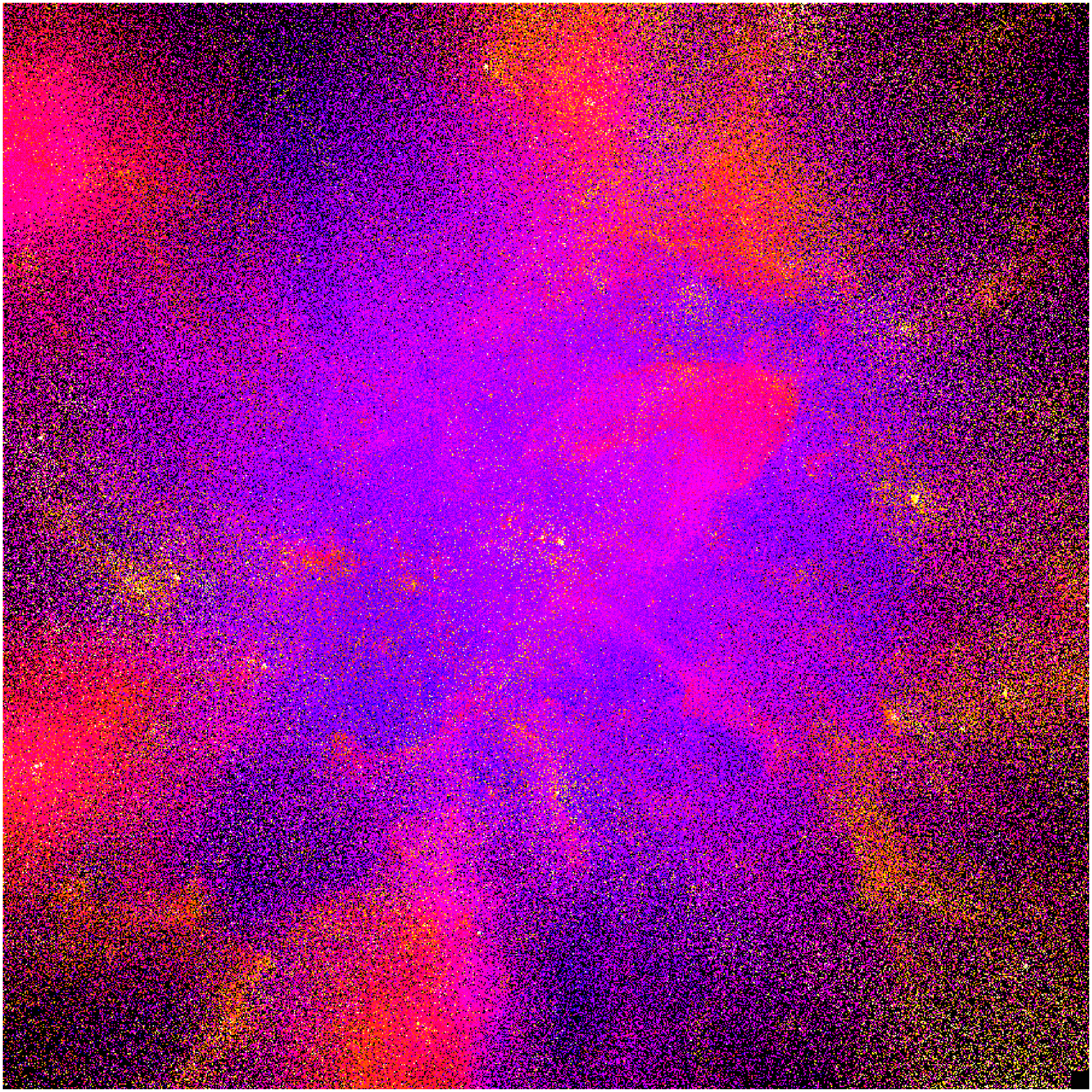}}
}
\caption{Temperature maps centered on a massive cluster having
  $M_{200}\simeq 1.3\times 10^{15}\msun$. From left to right, panels
  show maps for the {\tt \nr}, {\tt \w} and {\tt \agn}
  simulations. Each panel encompasses a physical scale of $6.25\hm$ a
  side. Colder regions are shown with brighter colours.}
\label{fig:tmaps}
\end{figure*}
In order to show the different degree of thermal complexity of the ICM
generated by the different models, we show in Fig.~\ref{fig:tmaps}
the temperature maps of a massive cluster for the {\tt \nr},
{\tt \w} and {\tt \agn} cases.  Quite apparently, in the {\tt \nr}
simulations there is more gas in relatively small cold clumps that
bias \tsl\  low and contribute to the increase in scatter\footnote{We note that the 
coldest gas particles,  those with temperature below $0.3\,keV$, are not taken into 
account in the computation of \tsl\ .}. 
A smaller
amount of cold gas is instead present in the radiative
simulations. For the {\tt \w} simulations, this is mostly due to gas
removal by overcooling. As a result, only relatively hot gas, having
longer cooling time, is present in substructures. In this case, a
smaller amount of cold gas characterises not only the sub-clumps, but
also the ram--pressure stripped gas that forms the comet-like
features associated with merging events. As for the {\tt \agn} simulation,
in this case relatively cold gas is removed from sub-clumps by a
twofold effect of BH feedback. Firstly, this feedback heats gas 
before a substructure is merged into the main cluster halo. Secondly,
the  more diffuse gas distribution makes sub-halos more prone to be
ram--pressure stripped as they move through the hot pressurised
atmosphere of the cluster.
The smoother temperature distribution in the radiative simulations
increases \tsl\, at fixed mass, with respect to the non-radiative
case, and decreases the scatter due to the presence of cold clumps,
thus bringing the simulated mass--temperature relation in better
agreement with observations. Quite interestingly, the effect of AGN
feedback in halos below $10^{14}\msun$ is that of slightly increasing
$T_{sl}$ with respect to the {\tt \w} simulations, thus producing a
shallower slope of this scaling relation, in better agreement with
observations.  Despite the different manner in which cooling/heating
processes remove gas from subclumps in each case, the \tslm\ relations
obtained in both of our radiative runs have similar normalisations and
slopes.  The fact that the slope we obtain for this relation is pretty
similar in all cases ($\alpha \sim 0.5$) indicates that the \tslm\
relation is weakly sensitive to baryonic physics
\citep[e.g.,][]{short_2010}.  However, we see that none of our
simulations reproduce the self-similar scaling for the \tslm\
relation.

\begin{figure*}
{\includegraphics[width=8cm]{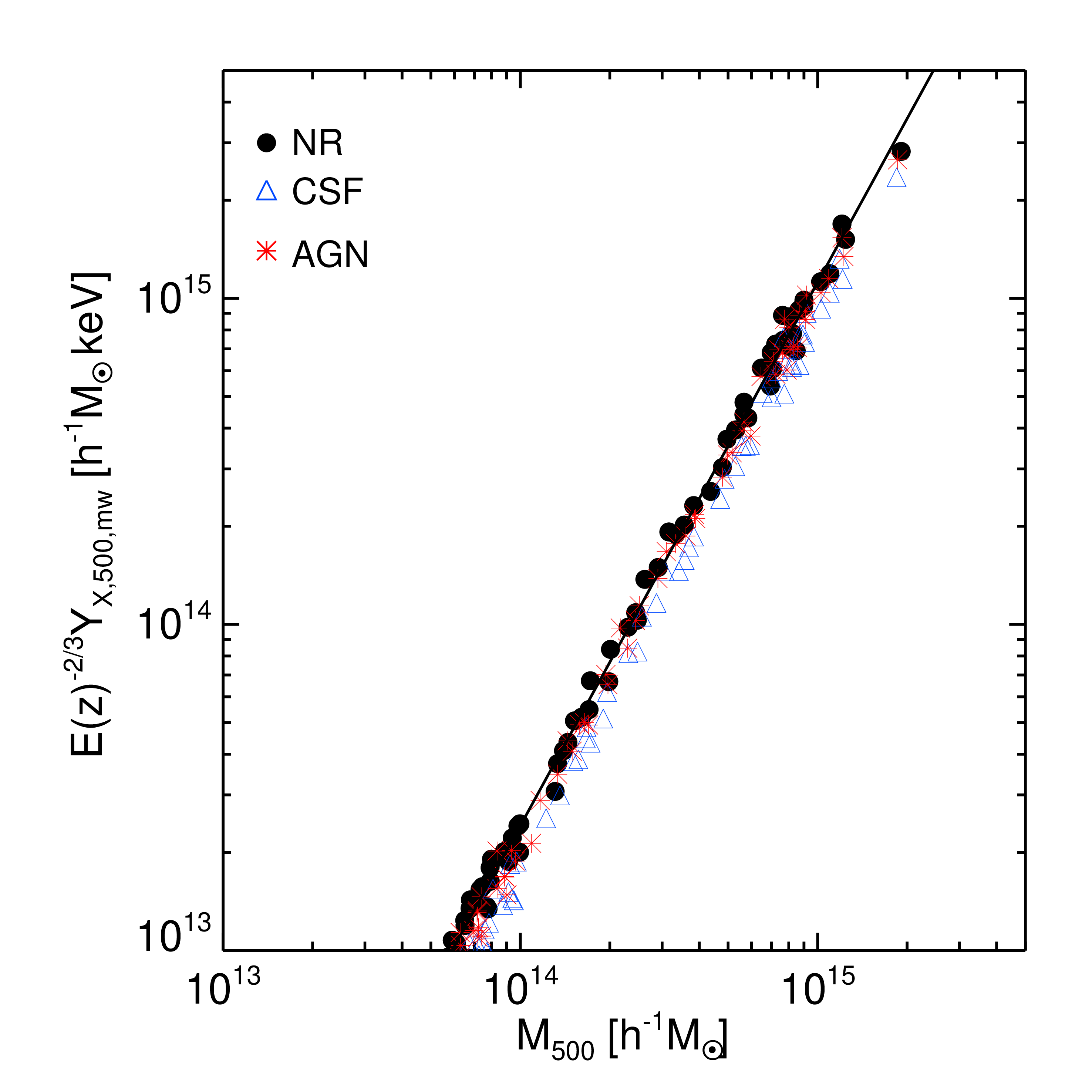}}
{\includegraphics[width=8cm]{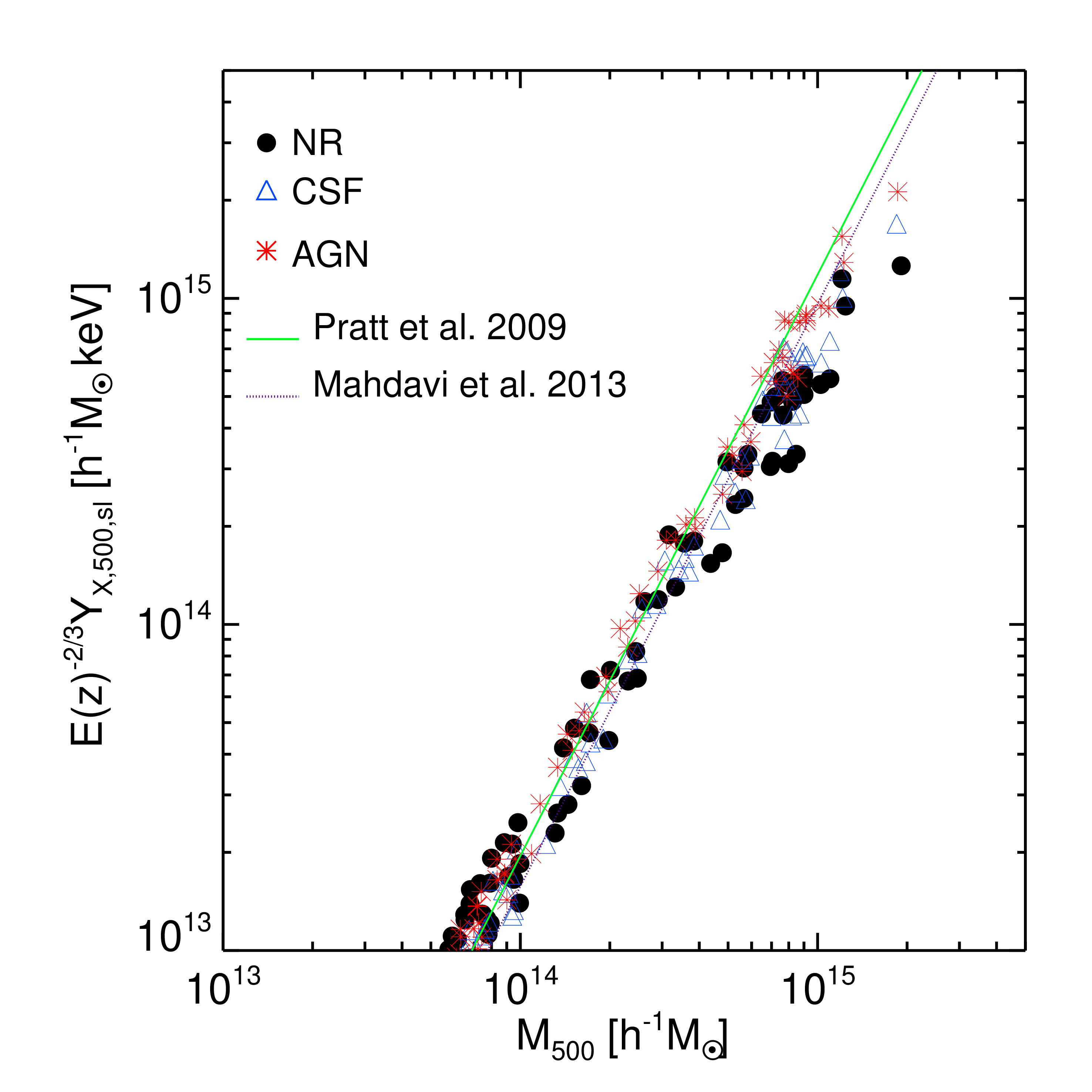}}
\caption{Relation between $Y_X$ and total mass within $R_{500}$ at
  $z=0$.  In the left panel, our results on the \yxmwm\ relation are
  compared with the self-similar scaling (black continuous line).  In
  the right panel we compare our results on the \yxmsl\ relation with
  the observational results from \citet{Pratt_2009} and
  \citet{Mahdavi_2013}.  In both panels, results for our sample of
  clusters within the {\tt \nr}, {\tt \w}, and {\tt \agn} runs as
  represented by black circles, blue triangles and red stars,
  respectively.}
\label{fig:yx_m}
\end{figure*}

As for the \yxm\ relation, we remind  that both simulations
\citep[e.g.,][]{Kravtsov_2006,Stanek_2010,short_2010, Fabjan_2011} and
observations \citep[e.g.,][]{Arnaud2007, Maughan_2007, Vikhlinin_2009}
indicate that \yx\ is a low-scatter mass proxy, even in the presence
of significant dynamical activity. Furthermore, X-ray observations have
shown that the measured slope for this relation agrees with the
self-similar prediction $\alpha=5/3$ (see Eq.~\ref{eq:yx_m}).

Figure \ref{fig:yx_m} shows the local \yxm\ relation obtained for our
sample of clusters within the {\tt \nr}, {\tt \w}, and {\tt \agn}
simulations.  In the left panel, we show our results on the \yxmwm\
relation and compare it with the self-similar scaling. Based on this
plot, we confirm that the total thermal content of the ICM is tightly
connected to cluster mass in a way that is weakly sensitive to the
inclusion of different physical processes affecting the evolution of
the intra-cluster plasma \citep[e.g.][and references
therein]{Kravtsov_2006,short_2010,Fabjan_2011,Kay_2012}. Residual
variations for the {\tt \w} simulations with respect to the {\tt \nr}
ones are due to the removal of gas from the hot phase as a consequence of
overcooling, which causes a decrease of $Y_X$ at fixed mass. The
mass-dependent efficiency of cooling in this case causes a small
deviation from the self-similar slope of the {\tt \nr}
simulations. Conversely, including AGN feedback has the effect of
partially {\it preventing} gas removal from cooling, thus slightly
increasing the normalisation of the \yxmwm\ relation. 

$Y_{X,mw}$ is a key physical quantity since it is the X-ray analogue
of the integrated Compton parameter $y$, a measure of the gas pressure 
integrated along the line--of--sight  \citep[e.g.,][]{Kravtsov_2006}. 
The total SZ signal, integrated over the cluster 
extent, is proportional to the integrated Compton parameter $Y_{SZ}$, which relates 
to $Y_X$ as $Y_{SZ}D_A\propto Y_X$, where $D_A$ is the angular distance to the system.
Therefore, understanding the scaling and evolution of $Y_{X,mw}$ is important not only as a probe 
of the ICM physics, but also to exploit the combination of X-ray and SZ data 
\citep[e.g.,][]{Arnaud_2010}.

In the right panel of Fig.~\ref{fig:yx_m}, our results for the \yxmsl\ relation are compared
with the observational relation from \citet{Pratt_2009} and from
\citet{Mahdavi_2013}. We remind the reader here that \citet{Pratt_2009} estimated
masses from the application of hydrostatic equilibrium to X-ray data,
while \citet{Mahdavi_2013} measured cluster masses from weak lensing
data.
The \yxmsl\ relation obtained from the {\tt \nr} simulations is
shallower than the observed relations. The increase of $T_{sl}$ when
passing from the {\tt \nr} to the {\tt \w} and the {\tt \agn} sets
also causes a corresponding progressive increase of $Y_X$, thus
improving the agreement with observational results.

The fact that the {\tt \w} and {\tt \agn} runs yield \yxm\ relations
that are close to each other and also close to the self-similar
prediction implies that \yx\ must be relatively unaffected by the
non-gravitational heating in these models. In the case of the 
{\tt \agn} run, this  arises because AGN feedback removes gas from the
central regions of halos, reducing the gas mass within $R_{500}$, but
this is offset by an increase in gas temperature caused by the
continuous injection of energy from the BH feedback.

\subsubsection{$L_{X}-T$ relation}

\begin{figure} 
{\includegraphics[width=8cm]{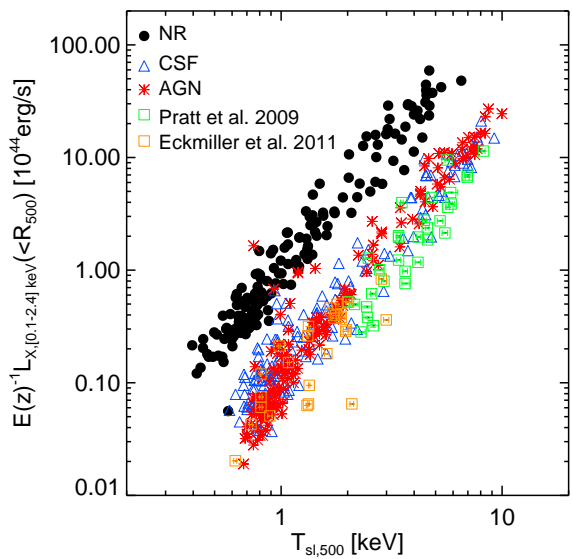}} 
\caption{X-ray luminosity, computed within the $[0.1-2.4]\,$keV energy
  band, as a function of spectroscopic-like temperature within
  $R_{500}$ (excluding the central regions,
$r<0.15R_{500}$) at $z=0$. Results for our sample of clusters within the
  {\tt \nr}, {\tt \w}, and {\tt \agn} runs are represented by black
  circles, blue triangles and red stars, respectively.  Observational
  results from \citet{Pratt_2009} and \citet{Eckmiller_2011} are used
  for comparison.}
\label{fig:l_t}
\end{figure}

Figure \ref{fig:l_t} shows the $L_X$--$T_{sl}$ relation for the sample of
clusters in our {\tt \nr}, {\tt \w}, and {\tt \agn} sets. Here X-ray
luminosities are computed in the [0.1--2.4] keV energy band.  We
compare with observational data on the scale of galaxy clusters and
groups from \cite{Pratt_2009} and \cite{Eckmiller_2011}, respectively.
In this case there is no special reason to compare our results to predictions
of the self-similar model. In fact, violation of self-similarity is
expected as a consequence of computing $L_X$ in a specific energy band 
and using $T_{sl}$ instead of $T_{mw}$.
%

\begin{figure*}
{\includegraphics[width=8cm]{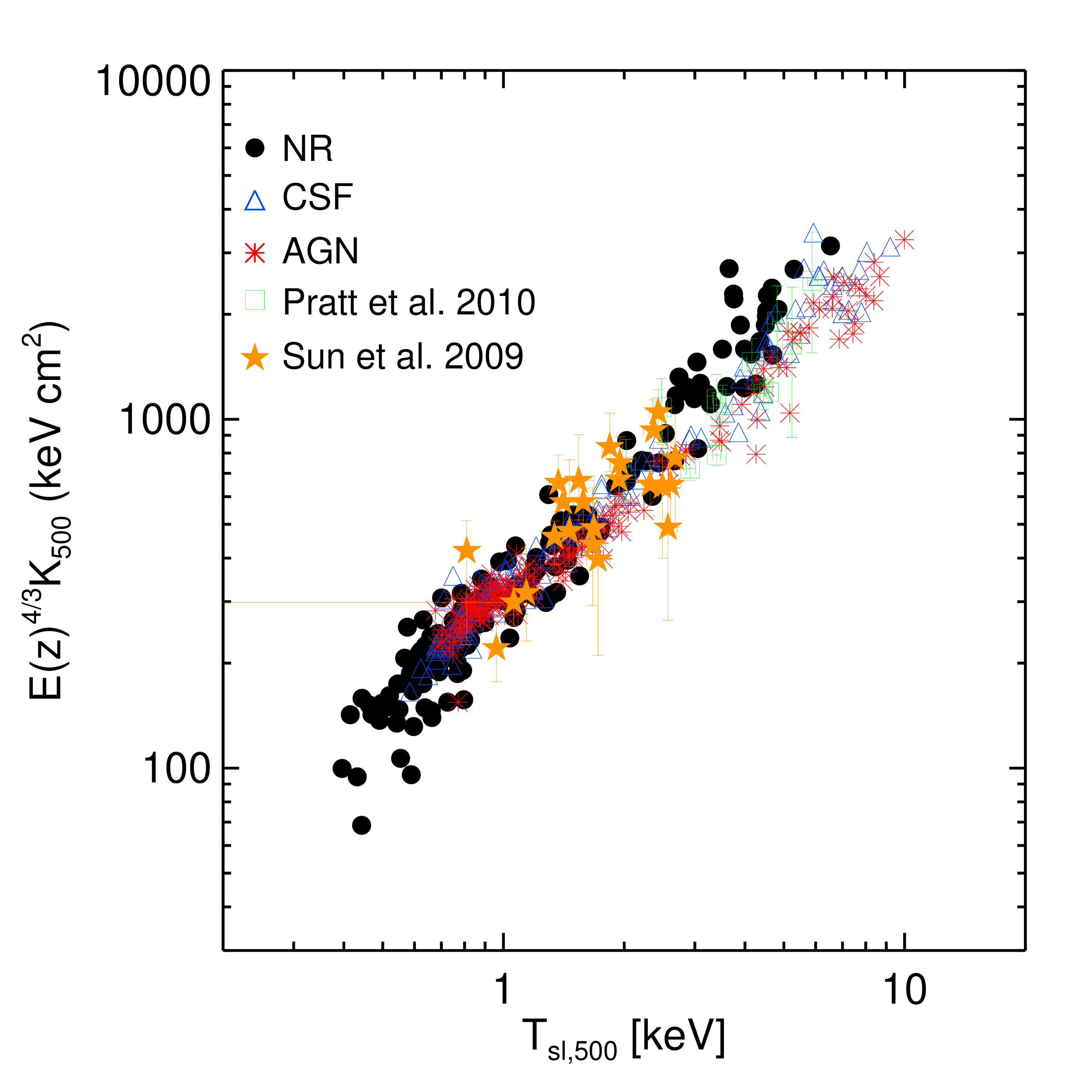}}
{\includegraphics[width=8cm]{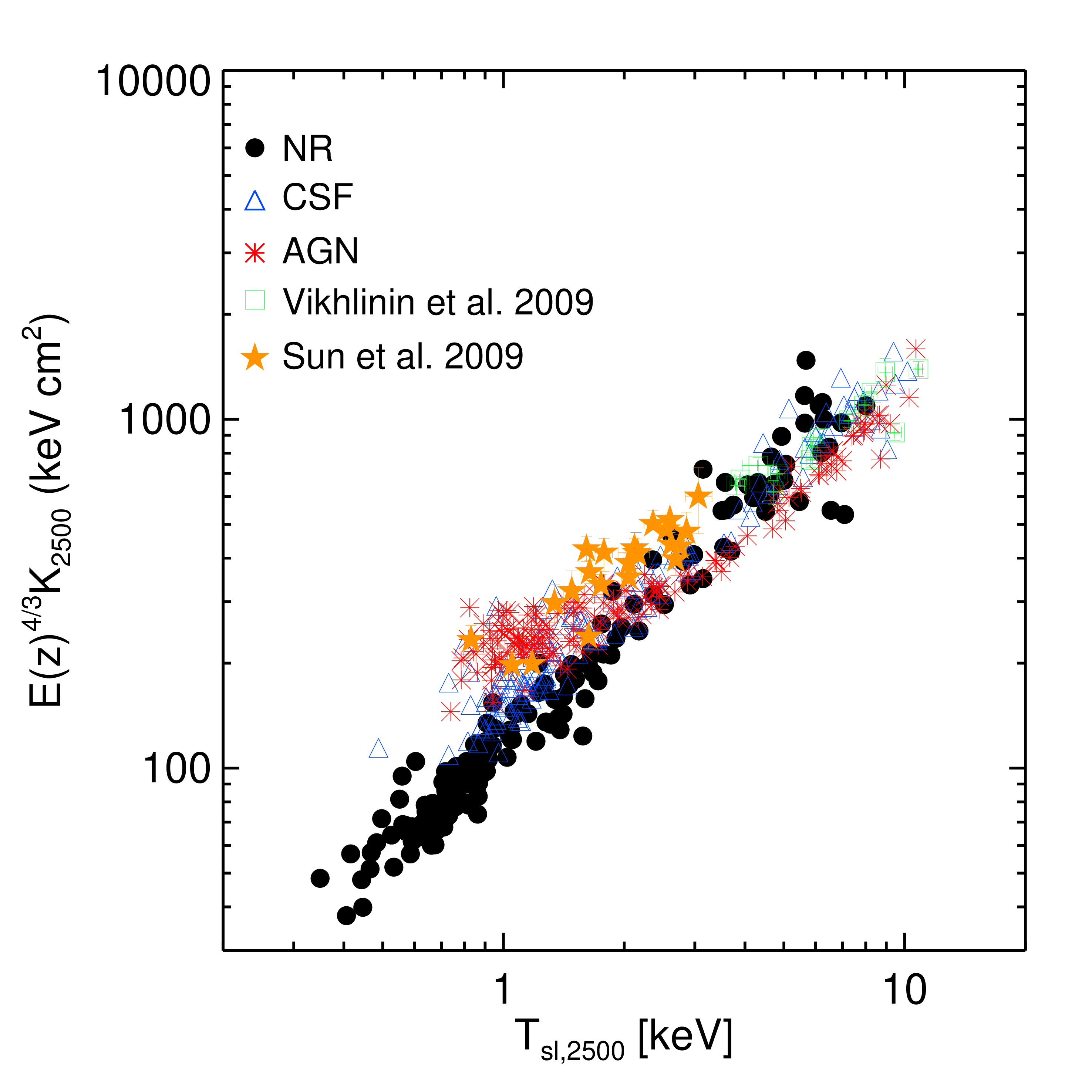}}
\caption{Entropy as a function of spectroscopic-like temperature
  at $R_{500}$ (left panel) and at $R_{2500}$ (right panel) at
  $z=0$.  Results for our simulated clusters within the {\tt \nr},
  {\tt \w}, and {\tt \agn} sets are shown as black circles, blue
  triangles and red stars, respectively.  Observational results from
  \citet{Pratt_2010}, \citet{Sun2009} and \citet{Vikhlinin_2009} are
  used for comparison.}
\label{fig:s_t}
\end{figure*}

We note that the {\tt \nr} runs fail to reproduce the observed
$L_{X}-T_{sl}$ relation and produce clusters that are more luminous
than the observed ones.  On the contrary, both of our radiative runs
produce a significant reduction of X-ray luminosity at all scales
obtaining, therefore, results that are closer to the observational
data.  In the {\tt \w} simulations the reduction of X-ray luminosity
is the consequence of overcooling, which causes an exceedingly high
removal of hot gas from the X-ray emitting phase and forces a too large
fraction of gas to be converted into stars \citep[e.g.,][and references
therein]{Kravtsov2005, fabjan_etal10, Puchwein2010, sembolini_etal12,
  Planelles_2012}.  Quite interestingly, both the {\tt \w} and {\tt
  \agn} models yield almost identical \lxtslband\ relations, despite
the fact that the latter reduces the amount of stars to levels
consistent with observational results (see Fig.~2 from
\citealt{Planelles_2012}; see also \citealt{Puchwein2010,
  Battaglia2012}).  In this case, gas removal by the action of AGN
feedback compensates the larger amount of gas left in the diffuse
phase by the reduction of star formation. Furthermore, AGN feedback is
slightly more efficient in decreasing X-ray luminosity at the scale of
galaxy groups, as a consequence of the more efficient removal of gas
in less massive systems, with shallower potential wells. This turns
into a steepening of the $L_X$--$T_{sl}$ relation with respect to the
{\tt \w} case, thereby recovering the observational results better.

If we use instead the values of $T_{sl}$ computed without
  excising the core regions, we would obtain similar $L_{X}-T_{sl}$
  relations, in terms of normalisation and slope, for the three set of
  simulations.  However, some low--mass systems show important
  deviations in their temperatures, contributing to increase the
  scatter around the mean relation, especially within the {\tt \agn}
  simulations.

Our results are in line with previous results from simulations
including different implementations of AGN feedback
\citep[e.g.,][]{Puchwein2008, short_2010, fabjan_etal10}.  It is
important to note that the AGN feedback implemented in these works
differs among each other both in the implementation of the AGN feedback
mechanism and in the treatment of the metallicity dependence of the
cooling function.  However, in all cases, results highlight the fact
that, almost independently of the details of the heating mechanism,
feedback energy associated with accretion onto SMBH is indeed able to
reproduce a realistic $L_X$--$T$ relation.

\subsubsection{$K-T$ relation}
We show in Fig.~\ref{fig:s_t} the relation between the entropy and
the spectroscopic-like temperature at $R_{500}$ (left panel) and
$R_{2500}$ (right panel) for our sample of simulated clusters in the
{\tt \nr}, {\tt \w}, and {\tt \agn} sets.  We compare our results with
observational data for clusters and groups from \citet{Pratt_2010},
\citet{Sun2009} 
and \citet{Vikhlinin_2009}.

At $R_{500}$ we note that all simulation sets produce an
entropy--temperature relation which is in good agreement with
observational results at the group scale, with radiative simulations
being more consistent with observations of massive
clusters. However, this similarity of the scaling relation does not
imply that different feedback mechanisms leave entropy unaffected at
$R_{500}$ within groups. In fact, as discussed above, the value of
$T_{sl}$ at fixed mass increases for the radiative simulations, an
effect which is more pronounced for the {\tt \agn} case. As we shall
discuss in Section 3.3.2, the increase of entropy in the {\tt \w} case
is due to the combination of two effects: selective removal by cooling
of low-entropy gas, which has shorter cooling time; and inflow of higher
entropy gas from outer cluster regions caused by the lack of pressure
support associated with overcooling. As in the {\tt \agn} case, the
increase of entropy in groups is instead due to gas heating
that in fact prevents the excess of cooling. The corresponding
increase of entropy in radiative simulations of groups causes a shift
of all the points along the same direction traced by the $K$--$T$
relation of the non-radiative runs. In massive systems, different
feedback mechanisms have a small impact on the entropy level at
$R_{500}$, so that the effect on the $K$--$T$ relation is only induced
by the variation of $T_{sl}$ produced by the presence of cooling and of
different feedback mechanisms.

As expected, the effects of the different ICM physics are more
pronounced at $R_{2500}$ (right panel of Fig.~\ref{fig:s_t}).  At the
scale of small clusters and groups, AGN feedback provides a
significant increase of the entropy level, which causes a significant
deviation with respect to the prediction of non-radiative simulations
and better reproduces observational results on the entropy excess for
systems with $T_{sl}\simeq 1$ keV.

\subsection{Profiles of ICM properties}

\begin{figure*}
\centerline{\includegraphics[width=18cm]{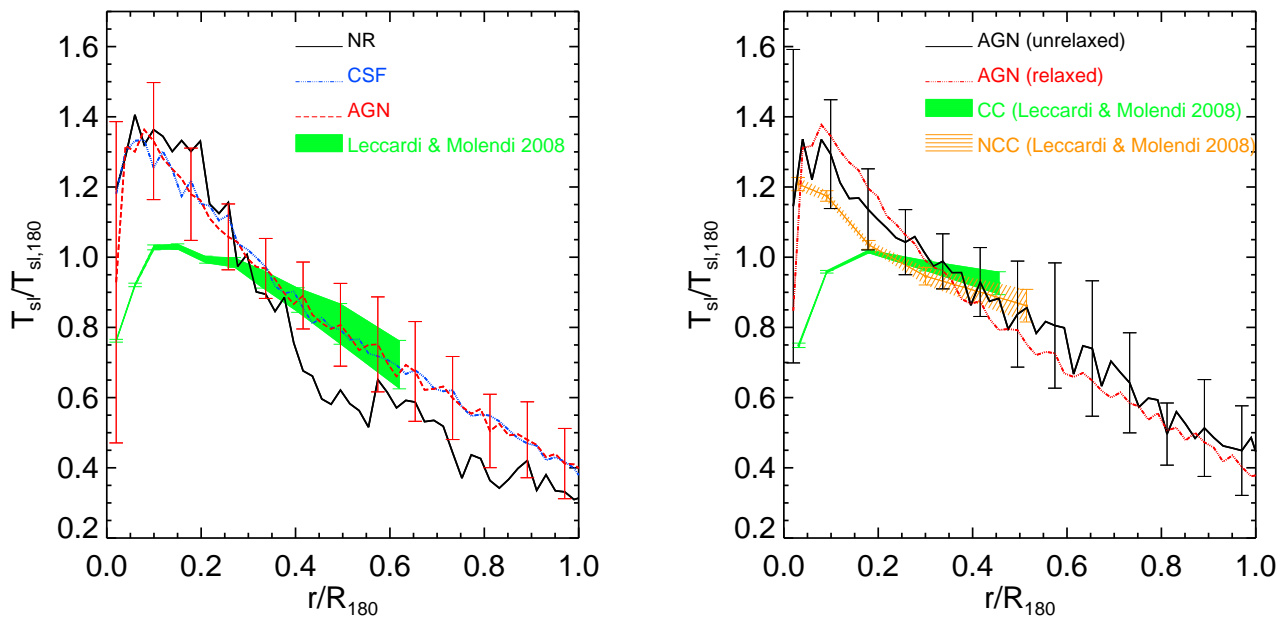}}
\caption{{\it Left panel:} mean $T_{sl}/T_{sl,180}$ radial profiles
  out to $R_{180}$ at $z=0$ for our sample of massive clusters with
  $T_{sl,500}>3$ keV.  Black continuous, blue dotted-dashed and red
  dashed lines stand for the mean profiles in the {\tt \nr}, {\tt \w}
  and {\tt \agn} simulations.  For the sake of clarity, error bars
  showing $1\sigma$ scatter are shown only for the {\tt \agn} case.
  {\it Right panel:} mean $T_{sl}/T_{sl,180}$ radial profiles computed
  separately for the relaxed/unrelaxed cluster subsamples in our {\tt
    \agn} run.  Black continuous and red dotted-dashed lines stand for
  the mean profiles of unrelaxed and relaxed systems, respectively,
  with the $1\sigma$ scatter only shown for the unrelaxed systems.
  In both panels, we compare our results with the observed temperature
  profiles from \citet{Leccardi_2008b}, which are represented by the
  coloured shadowy areas.  }
\label{fig:profiles_t}
\end{figure*}

Radial profiles of the ICM thermal properties are more sensitive than
scaling relations to the precise manner in which cooling, star
formation and feedback processes are described in numerical
simulations. In this Section we examine whether our different feedback
schemes are able to reproduce the temperature, entropy and pressure
profiles of observed local clusters.

Owing to the reference observational results that we will consider, in
the following we will restrict the analysis of these radial profiles
only to relatively massive clusters in our simulation sets, with
$T_{sl,500}\gtrsim 3\, keV$.  Within our simulations we identify about 36 
such objects at $z=0$.

In order to examine the effect of the dynamical state of the clusters
on the radial profiles, we divide our sample into dynamically relaxed
and unrelaxed systems.  We perform this classification by simply
measuring the offset between the position of the potential minimum of
the cluster and the centre of mass of all the particles within its
virial radius \citep[e.g.,][]{Crone_1996, Thomas_1998, Power_2012}. If
this offset is larger (smaller) than $0.07$ (in units of the cluster
virial radius), then the cluster is considered to be dynamically
unrelaxed (relaxed).  With this classification, the fraction of
relaxed systems found in our sample is $\sim
80$ per cent of the whole sample within each set of simulations.

\subsubsection{Temperature profiles}

The left panel of Fig.~\ref{fig:profiles_t} shows the average
spectroscopic-like temperature profile of clusters for our {\tt \nr},
{\tt \w} and {\tt \agn} simulations out to $R_{180}$.  All profiles
have been normalised to the characteristic mean cluster temperature
within $R_{180}$, $T_{sl,180}$.  The mean temperature profile obtained
from \cite{Leccardi_2008b} is shown for comparison as a green 
region.  \cite{Leccardi_2008b} measured radial temperature profiles
for a sample of $\approx 50$ hot galaxy clusters, selected from the
\xmm\ archive.  Most of the clusters in this sample ($\approx 2/3$)
belong to the REFLEX Cluster Survey catalog \citep{Bohringer_2004}, a
statistically complete X-ray flux-limited sample of 447 galaxy
clusters, and a dozen objects belong to the \xmm\ Legacy Project
sample \citep{Pratt_2007}, which is representative of an X-ray
flux-limited sample with $z<0.2$ and $\mathrm{k}T>2$~keV. Similarly to
other X-ray measurements of temperature profiles
\citep[e.g.,][]{Sanderson_2006, Vikhlinin2006, Pratt_2007,
  Arnaud_2010}, the temperature peaks at $r\mincir
0.2\,R_{180}$, with a gentle decline at small radii, which is the
signature for the presence of cool cores.

As already demonstrated by previous analysis of cluster simulations
\citep[e.g.,][]{Loken_2002, Borgani_2004,
Kay_2007,Pratt_2007,Nagai_2007_2, fabjan_etal10, short_2010,
Vazza_2010}, almost independently of the physical processes
included, the temperature profiles for our sample of galaxy clusters
have a slope that agrees quite well with observations in outer cluster
regions, $r\gtrsim 0.2R_{180}$, where the effect of cooling and
feedback is relatively unimportant. However, at such radii we note
that the temperature profile for the {\tt \nr} simulations is
systematically lower than for the radiative cases, and with a rather
irregular behaviour. As already discussed in Section 3.2.1, this is
a consequence of the spectroscopic-like estimate of the
temperature, which gives more weight to the colder gas component associated with
substructures, which are more prominent at relatively large radii (see
also Fig.~\ref{fig:tmaps}). Therefore, the wiggles in the {\tt \nr} profile are
just due to the effects of substructures, that persist even after
averaging over the set of simulated clusters.

As for the core regions, we see that in all cases simulated
temperature profiles are higher than the observed ones. This
discrepancy persists also in the presence of radiative cooling.  As
discussed above, this is a paradoxical effect of cooling due to the
adiabatic compression of gas flowing in from cluster outskirts to
compensate for the lack of pressure support caused by too much gas
cooling out of the hot phase. To prevent this overcooling and reduce
the central values of the temperature, one should require that a
suitable feedback mechanism keeps the gas at an intermediate
temperature. Since this ``cool'' gas formally has a short cooling time,
preventing it from cooling out of the X-ray emitting phase requires
heating from energy feedback to exactly balance radiative losses. The
comparison of our results with observational data in the right panel
of Fig. \ref{fig:profiles_t} shows that even AGN feedback is not
effective in creating the correct structure of ``cool cores'' in
massive galaxy clusters. 
 
To look for differences between clusters in different dynamical
states, the right panel of Fig.~\ref{fig:profiles_t} shows the mean
temperature profiles computed separately for the relaxed and unrelaxed
cluster subsamples in our {\tt \agn} set.  Now we compare our results
with the observed samples of cool core (CC) and non-cool core (NCC)
clusters from \citet{Leccardi_2008b}.  These authors classified
clusters as CC if the central temperature is significantly lower than
$T_\mathrm{180}$, while morphologically disturbed or NCC systems are
those for which the temperature profile does not significantly
decrease. Admittedly, this classification is based on a criterion that
is different from that we followed to classify clusters as relaxed and
unrelaxed. Still, it is quite interesting to note that temperature
profiles of our simulated clusters do not depend on their dynamical
state. Our results extend to the AGN feedback case a similar result
found by \cite{Eckert_2013_1} for the non-radiative AMR simulations
by \cite{Vazza_2010}.  In general, profiles from simulations tend to
agree better with results from NCC systems, which also display rather
steep temperature gradients down to small radii. This result is
consistent with the expectation that the adopted model of AGN feedback
is not capable of producing the heating/cooling balance, which is
responsible for the stabilisation of cool cores in relaxed systems.

 Our results, which are generally in line with independent analyses of
  simulations including different versions of AGN thermal feedback,
  are not able to convincingly reproduce the observed thermal
  properties of cluster core regions \citep[e.g.,][]{Kravtsov_2012}.
  The reason for these discrepancies may be related to both the
  limited numerical resolution achievable with cosmological
  simulations, and the difficulty of providing a coherent description
  of the complex interplay between AGN feedback and a number of other
  physical processes (e.g., turbulence ICM motions, non-thermal
  pressure support, magnetic fields).

  In addition, observational results indicate that sub-relativistic
  jets from the BH hosted in central cluster galaxies shocks the
  surrounding ICM, thereby producing bubbles of high-entropy gas.  In
  this regard, recent numerical experiments \citep[e.g.][]{Omma_2004,
    Gaspari_2011, Barai_2013} indicate that the inclusion of
  mechanical AGN feedback in cosmological simulations seems to be an
  improvement to be implemented \citep[e.g.][]{Martizzi_2012}, along
  with the exploration of accretion models different from the standard
  Bondi criterion.

\subsubsection{Entropy profiles}
\begin{figure*}
\centerline{\includegraphics[width=17cm]{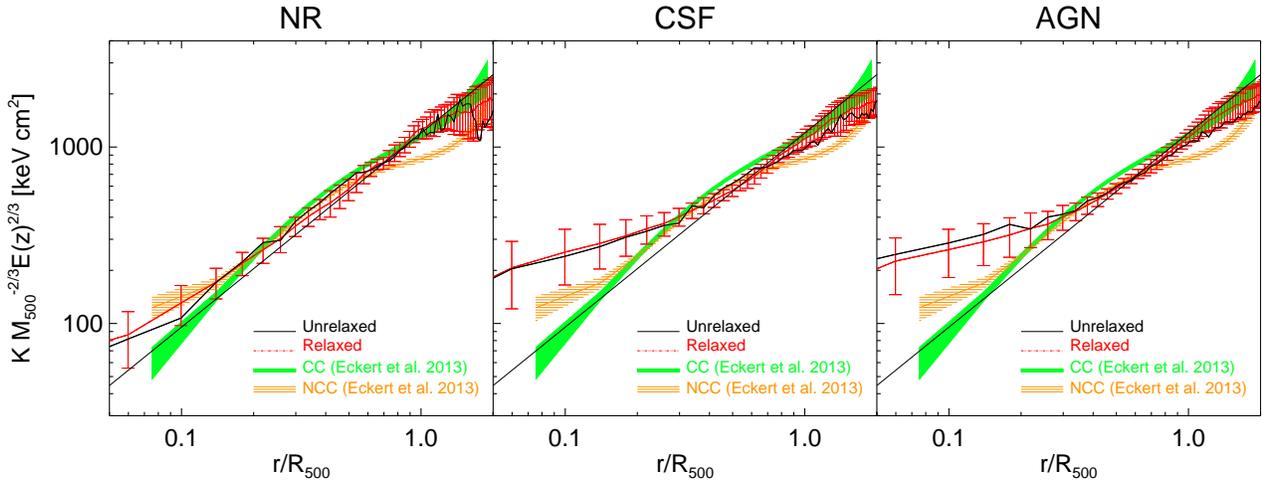}}
\vspace{0.5cm}
\caption{Mean entropy radial profiles at $z=0$ for the sample of
  massive clusters in our {\tt \nr} (left panel), {\tt \w} (middle
  panel) and {\tt \agn} (right panel) simulations.  Mean profiles are
  computed separately for the unrelaxed and relaxed subsamples of
  simulated clusters (black continuous and red dotted-dashed lines,
  respectively).  Error bars correspond to the $1\sigma$ scatter
  around the mean profile.  We compare our results with the observed
  cluster entropy profiles by \citet{Eckert_2013_1}, which are
  represented by the coloured areas indicating the average scaled
  profile $\pm1\sigma$ dispersion around it.  The thinner solid black line in
  all panels shows for reference the $K\propto r^{1.1}$ profile.}
\label{fig:profiles_s}
\end{figure*}

The entropy distribution of the ICM has long been a
crucial diagnostic to study the impact of non-gravitational
processes, related to galaxy formation, on the diffuse cosmic
baryons. In fact, entropy preserves a record of the physical processes
that determine the thermal history of the ICM \citep[e.g.,][for a
review]{Voit_2005}.

Analytical models based on spherical collapse predict that, if shock
heating is the only mechanism acting to raise the entropy of the gas,
entropy scales with radius as $K\propto r^{1.1}$ outside of central
cluster regions \citep[e.g.,][]{Tozzi_2001}.  Cosmological simulations
that only include gravitational heating confirm this prediction
although give rise to slightly steeper entropy profiles in cluster
outskirts, with $K\propto r^{1.2}$ \citep[e.g.,][]{Voit_etal_2005,
  Nagai_2007_2, Planelles_2009}. These results from simulations are
generally in line with observational results
\citep[e.g.][]{Pratt_2010,Eckert_2013_1}.

At small radii, observed entropy profiles display a variety of
behaviours, depending on their dynamical state.  Indeed relaxed CC
systems show steadily decreasing profiles down to the smallest sampled
radii, while unrelaxed NCC clusters have entropy profiles that flatten
off in the core regions \citep[e.g.][]{Sanderson_2006, Vikhlinin2006, 
Pratt_2007, Arnaud_2010} depending on a number of factors such
as the temperature of the system and its dynamical state.  In
particular, hotter, more massive objects have a higher mean core
entropy \citep[e.g.,][]{Cavagnolo_2009,Sanderson_2009, Pratt_2010}.

Figure \ref{fig:profiles_s} shows the mean entropy radial profiles for
the subsamples of relaxed and unrelaxed clusters in the {\tt \nr},
{\tt \w} and {\tt \agn} runs.  We compare these mean entropy profiles
with the profiles from \cite{Eckert_2013_1}, who derived the
thermodynamic properties of the intra-cluster gas (i.e., temperature and
entropy) for a sample of 18 clusters by combining the SZ thermal
pressure from \planck\ and the X-ray gas density from \rosat.  Based
on the value of the central entropy \citep{Cavagnolo_2009}, six of
these clusters were classified as CC ($K_0<30$ keV cm$^2$), while the
remaining 12 are NCC.

Independently of the dynamical state of the systems, 
in outer cluster regions ($r\gtrsim 0.1 R_{500}$), the slope of the entropy
profiles of the simulated clusters in the {\tt \nr} set is consistent
with the observed ones (and close to the $K\propto r^{1.1}$ scaling),
although with a somewhat lower normalisation. 
At smaller radii there is more scatter both in observations and in
simulations. At these radii, the entropy profiles of simulated
clusters agree with those of the set of relaxed CC clusters analysed
by \cite{Eckert_2013_1}, independently of their dynamical state. 

As for the clusters in the {\tt \w} and {\tt \agn} simulations, both
relaxed and unrelaxed systems show entropy profiles that are also
broadly consistent in slope with the theoretical self-similar scaling
at large cluster-centric radii ($r\geq 0.3-0.4\ R_{500}$), thus
supporting the idea that gravity dominates the ICM thermodynamics in
outer cluster regions.  For inner regions the slope of the profiles in
simulations decreases and approaches that of the observed
profiles of NCC clusters.  Both in the {\tt \w} and in the {\tt \agn}
simulations the profiles for relaxed and unrelaxed systems are
virtually identical. Therefore, although introducing star formation
and feedback changes the entropy level in the central cluster regions,
such effects are unable to create the observed diversity between CC
and NCC systems. Furthermore, the similarity of the profiles obtained
for the {\tt \w} and {\tt \agn} simulations indicates that cooling is
responsible for setting the entropy level below which gas cools and
form stars, while the nature and efficiency of feedback determines
how much of this gas drops out of the hot phase.

\subsubsection{Pressure profiles}

\begin{figure}
\centerline{\includegraphics[width=9cm]{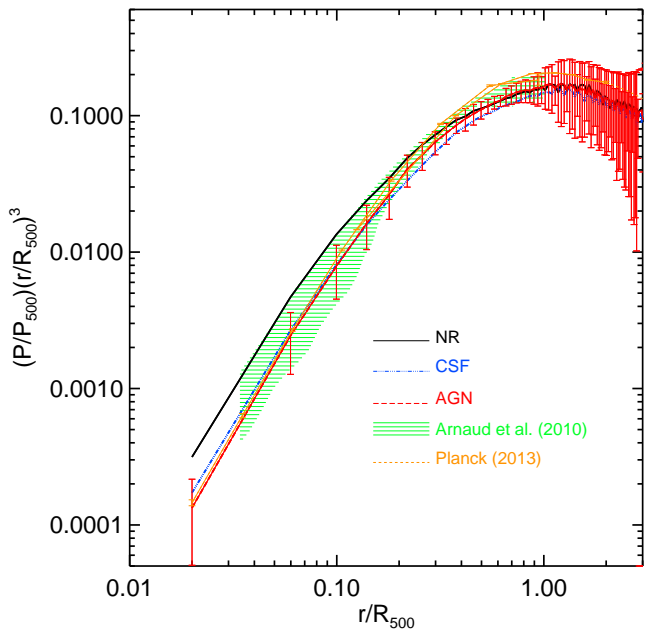}}
\caption{Mean pressure profiles (in units of $P_{500}$), for our sample
  of massive clusters with $T_{sl,500}>3$ keV.  In order to highlight
  the differences among the different physical models, these pressure
  profiles have been scaled by $(r/R_{500})^3$.  Black continuous,
  blue dotted-dashed and red dashed lines stand for the mean profiles
  in the {\tt \nr}, {\tt \w} and {\tt \agn} simulations, respectively.
  For the sake of clarity, error bars showing $1\sigma$ scatter are
  shown only for the {\tt \agn} simulations.  The green shadowy area
  corresponds to the average scaled profile $\pm1\sigma$ dispersion
  around it as observed by \citet{Arnaud_2010} from XMM observations,
  whereas the dashed orange line with error bars stands for the
  corresponding profile from SZ observations obtained by
  \citet{Planck_2013}.}
\label{fig:pres_prof}
\end{figure}

The analysis of the temperature and entropy profiles demonstrates that
clusters have a variety of behaviours in central regions, depending on
the presence and prominence of cool cores \citep[e.g.][]{Pratt_2010},
but outside of core regions they behave as a more homogeneous
population and follow the expectations of the self-similar model.  A
good illustration of the homogeneity of the ICM properties is
represented by the pressure profiles.

In Fig.~\ref{fig:pres_prof} we show the mean radial pressure profiles
obtained for the sample of clusters in our three sets of simulations.
The pressure profiles have been scaled by the `virial' pressure
$P_{500}$ as predicted by the hydrostatic equilibrium condition
\citep[see][]{Nagai_2007_2}:  
\begin{equation} 
P_{500} = 1.45\times 10^{-11} \,{\rm erg \; cm^{-3}}\; \left( \frac{M_{500}}{10^{15}\,h^{-1}M_{\odot}}\right)^{2/3} E(z)^{8/3} \ . 
\label{eq:p500} 
\end{equation}
In addition, in order to highlight the
differences among the different physical models, they have been scaled
as well by $(r/R_{500})^3$. With this scaling, the height of the
pressure profiles corresponds to the contribution per radial interval
to the total thermal energy content of the cluster
\citep{Battaglia_2012b}.

We compare our mean profiles with the X-ray observations of the
\rexcess\ sample by \cite{Arnaud_2010}, and with the SZ data from
\cite{Planck_2013}.  \cite{Planck_2013}, taking advantage of the
all-sky coverage and broad frequency range of the \planck\ satellite,
studied the SZ pressure profiles of 62 nearby massive clusters
detected at high significance in the 14-month nominal survey.  Most of
these clusters were individually detected at least out to $R_{500}$.
Then, by stacking the radial profiles, they statistically detected the
radial SZ signal out to $3R_{500}$.

From Fig.~\ref{fig:pres_prof} we see that the effect on pressure
profiles of radiative cooling, star formation and different forms of
feedback is generally relatively small, with all simulation models
agreeing rather well with observational results. To first order,
pressure profiles should just reflect the condition of hydrostatic
equilibrium within the potential wells that are established during the
cosmological assembly of clusters, thus following a nearly universal
profile \citep[e.g.][]{Nagai_2007_2,Arnaud_2010}. As such, they should
be relatively insensitive to the details of the thermodynamical status
of the ICM. This is the reason why mass proxies associated with
pressure, such as the integrated Compton--$y$ parameter or the 
aforementioned $Y_X$, are considered as robust mass proxies.
 
A closer look at Fig. \ref{fig:pres_prof} shows that radiative cooling
creates a decrease of pressure with respect to the non-radiative
case, at $r\mincir 0.2R_{500}$. As a result, the average pressure
profile for the {\tt \nr} case tends to be somewhat higher than the
observed one. At these small radii the two radiative simulation sets
produce results that are close to each other and in good agreement
with both X-ray \citep{Arnaud_2010} and SZ \citep{Planck_2013}
data. Our result on the weak sensitivity of central pressure on the
inclusion of AGN feedback apparently disagrees with the results
presented by \cite{Battaglia_2010}. Using simulations also based on
the SPH \gadget code, they showed instead that simulations with AGN
feedback produce pressure profiles that, in core regions, are below
those obtained without this feedback source. However, we note that,
besides the differences in the implementation of the AGN feedback
model, \cite{Battaglia_2010} carried out simulations within cosmological
boxes which are not large enough to include a significant population
of the hot systems, with $T_{sl,500}>3$ keV, considered in our
analysis. As a result, their pressure profiles give more weight to the
population of low-mass systems. In fact, we verified that restricting
our analysis only to clusters and groups with $T_{sl,500}<3$ keV, AGN
feedback has the effect of decreasing pressure in central
regions. Besides being in line with the findings of
\cite{Battaglia_2010}, this result also agrees with the expectation
that the total thermal content of the ICM is only weakly affected by
feedback sources in massive systems while being more sensitive to
feedback in systems with lower virial temperature.

A detailed comparison of the pressure profiles obtained in
observations and those derived from simulations including different
sets of physics would deserve a deeper study.  Since this is beyond
the scope of the present paper, we refer the reader to a future work
(Planelles et al. in prep.) in which we will present a detailed
analysis of the pressure profiles, along with a comparison with
observational data, and a study of their dependence on mass, evolution
with redshift and possible observational biases.

\section{Metal enrichment of the ICM}
\label{sec:metals}

The analysis of the content and distribution of metals within the
intra-cluster medium provides direct means of understanding the
interplay between the process of star formation, taking place on small
scales within galaxies, and the feedback and gas-dynamical processes
which determine the thermal properties of the ICM.  While star
formation affects the quantity of metals that are produced by
different stellar populations, various feedback and dynamical
processes affecting the gas (such as turbulent motions, ram-pressure
and tidal stripping) are responsible for displacing the metals from
star forming regions and determining their distribution in the ICM.

The model of chemical evolution included in our simulations
\citep{tornatore_etal07} allows us to follow the production of heavy
elements by accounting for the contributions from SN-II, SN-Ia and low
and intermediate mass stars. Whether or not these metals will then be
distributed throughout the ICM will depend on the competing roles of
cooling of enriched gas, which causes metals to get locked in
stars, and of feedback and gas-dynamical processes, which are
responsible for the circulation of metals outside star-forming
regions.  In this Section we compare predictions of the ICM metal
enrichment from the {\tt \w} and {\tt \agn} simulations to observational
data of galaxy clusters and groups.
We present results on the relation between global metallicity and
cluster mass, the Fe distribution and its corresponding abundance
profile, and the relative abundance of Si with respect to Fe. Finally,
we show how different feedback sources affect the relative metal
enrichment of the diffuse hot gas and of the stellar component.

We refer to observational results from \chandra, \xmm\ or \suzaku\ 
for comparison.  Before proceeding it is worth pointing out that
simulated metal abundances can only be expected to match the
observations to within a factor of $\sim 2$
\citep[see also][]{McCarthy_2010}, owing to the uncertainties in the
underlying chemical evolution model, e.g. related to the
adopted nucleosynthesis yields, the SN-Ia rates or the stellar lifetimes.

Throughout this Section, all metal abundances are scaled to the solar
abundances provided by \citet{Grevesse1998}.  In addition, unless
otherwise stated, we will rely on emission-weighted estimates of
metal abundances (see Eq.~\ref{eq:z_ew}). The emission-weighted
estimator has been shown to reproduce quite closely the values
obtained by fitting the X-ray spectra of simulated clusters, at least
for Fe and Si \citep[e.g.,][]{Kapferer_2007, Rasia_2008}, whereas
oxygen abundance may be significantly biased.  In the case of oxygen,
the difficulty of determining the continuum for the measurement of the
line width and the weakness of the corresponding lines at high
temperatures \citep[e.g.,][]{Rasia_2008} cause the emission-weighted
estimator to seriously overestimate the corresponding abundance,
especially for hot systems ($T\gtrsim 3\, keV$).

Since a unique extraction radius is not defined for the observed
catalogues, depending on the observational sample we compare with, we
adopt either $R_{500}$ or $R_{180}$ as common extraction radii to
compute metal abundances of simulated clusters. We verified that
adopting a larger extraction radius (e.g. $R_{vir}$) instead slightly
lowers the spectroscopic-like temperatures without changing
substantially the results on the abundance \citep[see as
well][]{Fabjan_2008}.

\subsection{The mass--metallicity relation}

\begin{figure*}
\centerline{\includegraphics[width=16cm]{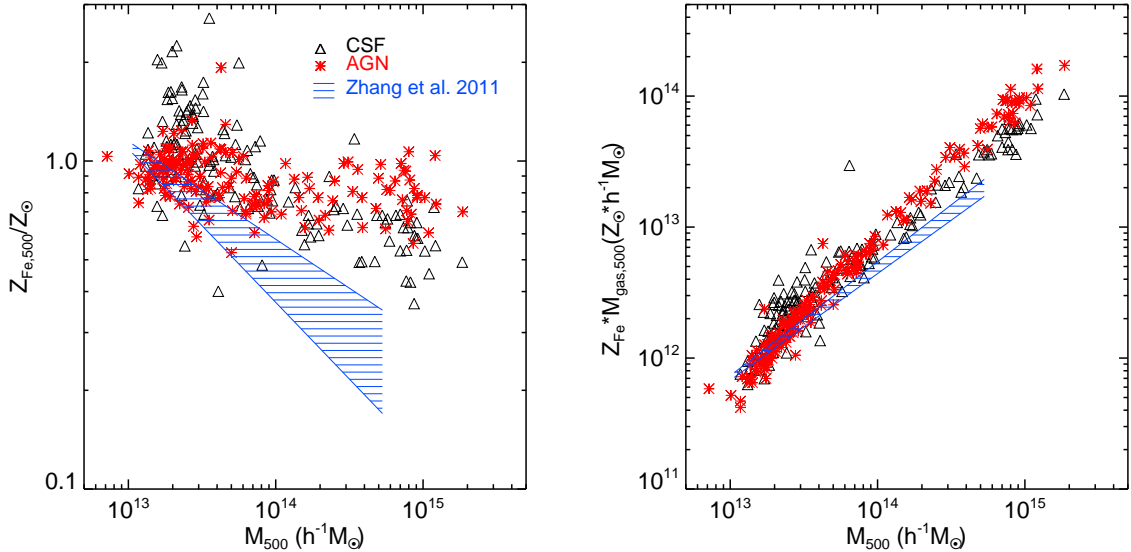}}
\caption{Emission-weighted iron abundance (left panel) and iron mass
  (right panel) within $R_{500}$ as a function of $M_{500}$.  Black
  triangles and red stars stand for the results from our {\tt \w} and
  {\tt \agn} runs, respectively.  Shaded blue area refers to the
  observational fitting obtained by \citet{Zhang2011}.}
\label{fig:iron_mass}
\end{figure*}

Figure~\ref{fig:iron_mass} shows, for the sample of clusters within
each of our radiative runs, the iron abundance (left panel) and iron
mass (right panel) as a function of the mass within $R_{500}$.  We
compare these results with the observational sample by
\citet{Zhang2011}, who investigated the baryon mass content for a
subsample of 19 clusters of galaxies extracted from the X-ray
flux-limited sample HIFLUGCS. 
In their analysis, ICM metallicity and gas mass are based on \xmm\ data,
while they derived cluster masses from measurements of the ``harmonic" 
velocity dispersion as described by \citet{Biviano2006b}.

The observational results by \citet{Zhang2011} indicate that less
massive clusters, which have lower gas-mass fractions but higher
iron-mass fractions, are more metal-rich.  A plausible explanation for
this result is that, in less massive galaxy clusters the
star-formation efficiency is higher, that is, more stars were formed
that have delivered more metals to enrich the hot gas.  On the other
hand, higher-mass clusters have a lower star-formation efficiency and
less metal enrichment in the hot gas by stars, in part because
feedback energy from, e.g., merging, is more efficient in quenching
star formation in their member galaxies.  In addition, given their
deeper potential wells, a larger amount of hot gas is accreted in more
massive clusters, diluting, therefore, the iron abundance more.

As we can infer from the analysis of Fig.~\ref{fig:iron_mass}, we
obtain a mild decrease of Fe abundance with increasing total cluster
mass (left panel), in agreement with the trends seen in previous
analyses of simulations including chemical enrichment
\citep[e.g,][]{Fabjan_2008,Dave_2008}.

As for the effect of feedback on this relation, we note that the
effect of including AGN is to weaken further the relation
between iron abundance and cluster mass. On the scale of groups,
$M_{500}\mincir10^{14}\msun$, AGN feedback decreases the iron
abundance with respect to the {\tt \w} case, to a level more
consistent with observations. For more massive systems,
$M_{500}\gtrsim10^{14}\msun$, AGN feedback has the opposite effect,
thereby producing values of iron abundance higher by a factor of $\sim
2$ than in real clusters.

These trends stem from the differential effect that AGN feedback has
on systems of different masses.  In low-mass systems enriched gas is
expelled at high redshift, thereby allowing metal poorer
gas to be later accreted from the surrounding IGM.  On the contrary,
in more massive systems enriched gas is more efficiently retained
within their potential wells. At the same time, suppression of star
formation due to the action of AGN feedback allows the metal
enriched gas to remain in the hot phase, instead of being locked in
stars, thus increasing the overall enrichment level of the ICM.
 
The right panel of Fig.~\ref{fig:iron_mass} also demonstrates that 
simulations with AGN feedback also provide the correct total iron mass
at the scale of groups. This implies that these simulations
provide both the correct amount of metals and the correct gas mass
within low-mass systems. In general, more massive systems have too
much iron mass. 

Generally speaking, this result shows that our simulations do not produce the
correct dependence of the ICM iron abundance on cluster mass. This is even
more true for the {\tt \agn} simulations, despite the fact that
including AGN feedback generally provides a closer agreement with the
observed thermal properties of the ICM (see Section 3). Ultimately, 
although our model of AGN feedback goes in the right
direction of reducing BCG masses, it is not yet efficient enough to
suppress star formation to the observed level at the centre of massive
clusters. The resulting BCGs are still too massive 
\citep[for a complete analysis see][]{RagoneFigueroa_2013}, 
and thus over--enrich the ICM to a too high level.

\subsection{Radial profiles of iron abundance}

The way in which metals are distributed in clusters carries information
both on the past history of star formation and on the feedback and
gas-dynamical processes which transport and diffuse them away from
galaxies.

\begin{figure}
\begin{center}
{\includegraphics[width=8.5cm]{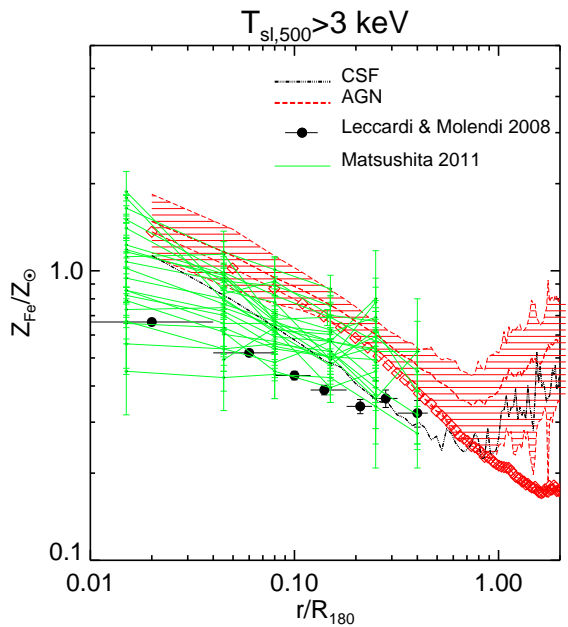}}
\end{center}
\caption{Mean emission-weighted iron abundance profiles for
  clusters with
  $T_{sl,500}\ge3$ keV in our radiative simulations.  Black
  dotted-dashed and red dashed lines stand for the {\tt \w} and {\tt
    \agn} models, respectively. Red
  dashed line connected by diamond symbols shows the corresponding
  mean mass-weighted iron abundance profile for the clusters in the
  {\tt \agn} set.  For the sake of clarity, error bars
  showing $1\sigma$ scatter from the mean profile are shown only
  for the {\tt \agn} run as a shaded area. Green error bars and filled
  circles with error bars show the observational results by
  \citet{Matsushita_2011} and \citet{Leccardi_2008}, respectively.}
\label{fig:metals_prof_ew}
\end{figure}

Figure \ref{fig:metals_prof_ew} shows the mean emission-weighted iron
abundance profiles obtained by averaging over the sample of hot
($T_{sl,500}\ge3\,keV$) systems in our radiative simulations out to
$2\times R_{180}$.  For completeness we overplot, only for the {\tt
\agn} run, the corresponding mean mass-weighted iron abundance
profile.  

We compare these with the observed
metallicity profiles by \citet{Leccardi_2008} and
\citet{Matsushita_2011}. For the former, we show the combined profiles obtained from their
analysis of about 50 clusters with $T \gtrsim 3$ keV, that were
selected from the \xmm\ archive in the redshift range $0.1 \leq z \leq
0.3$. \citet{Leccardi_2008} recovered metallicity profiles for these
systems out to $\simeq 0.4 R_{180}$. The results of this analysis show
a central peak of $Z_{Fe}$, followed by a decline out to $0.2
R_{180}$, while beyond that radius profiles are consistent with being
flat, with $Z_{Fe}\simeq 0.3 Z_{\odot}$, using the solar abundance
value by \citet{Grevesse1998}.  As for \citet{Matsushita_2011}, we
show individual Fe abundance profiles of the 24 nearby ($z < 0.08$)
clusters of galaxies observed with \xmm\ included in their sample, that
have an average ICM temperature above 3 keV.
The results obtained from this study are in qualitatively agreement with those found by
\citet{Leccardi_2008} although with a slightly higher normalization. 

The mean profile for the {\tt \agn} case is slightly shallower than in
the {\tt \w} case and with a higher normalization. The difference in
shape is consistent with the expectation that AGN feedback is
effective in redistributing metals within the ICM.  Our
mean profiles show the presence of abundance gradients in the central
regions whose shape is in reasonable agreement with observational
results. As discussed above, the higher enrichment level found in the
{\tt \agn} simulations is due to the efficiency with which this
feedback model is able to suppress star formation and displace metals
from star forming regions, thereby preventing highly-enriched gas with
short cooling time to be locked back into stars. 

In general, the fact
that simulations provide a shape of the metallicity profile which is
similar to the observed ones should be regarded as a remarkable
success of simulations.
Changes to specific parts of the chemical evolution model, e.g.
different sets of stellar yields \citep[e.g.,][]{tornatore_etal07,
 wiersma_etal09}; choice of the stellar IMF; a decrease in the
fraction of binary systems, which are the progenitors of SNe-Ia
\citep[e.g.,][]{Fabjan_2008}, could be implemented to reduce the overall
metal content, thus decreasing the normalization of the abundance
profiles to the observed level.

At large radii our simulations show a pronounced increase of the Fe
abundance up to the outermost radius with a relatively large scatter.
As we will discuss below, this increase of the emission-weighted iron
profile in outer cluster regions is due to the presence of halos
containing highly-enriched gas which has not yet been ram-pressure
stripped.  Being at high density, this gas provides a strong
contribution to the emission-weighted metallicity.
This interpretation is confirmed by the comparison with the
mass-weighted iron abundance profile which, instead, smoothly
decreases out to the outermost sampled radii. 

Our results in inner cluster regions are in general agreement with
those obtained from previous simulations including different forms of
energy feedback \citep[e.g.,][]{Bhattacharya_2008, fabjan_etal10,
  McCarthy_2010}.  However, none of these previous works show the
steep increase of the emission-weighted iron abundance profiles that
we find in outer cluster regions.  Nevertheless, we need to be careful
with this comparison since most of the results 
presented in the literature refer either to mass-weighted profiles or
to profiles only for low-temperature systems (below 3 keV), for which
we also see no such increase at large radii.

\begin{figure*}
\begin{center}
{\includegraphics[width=15.5cm]{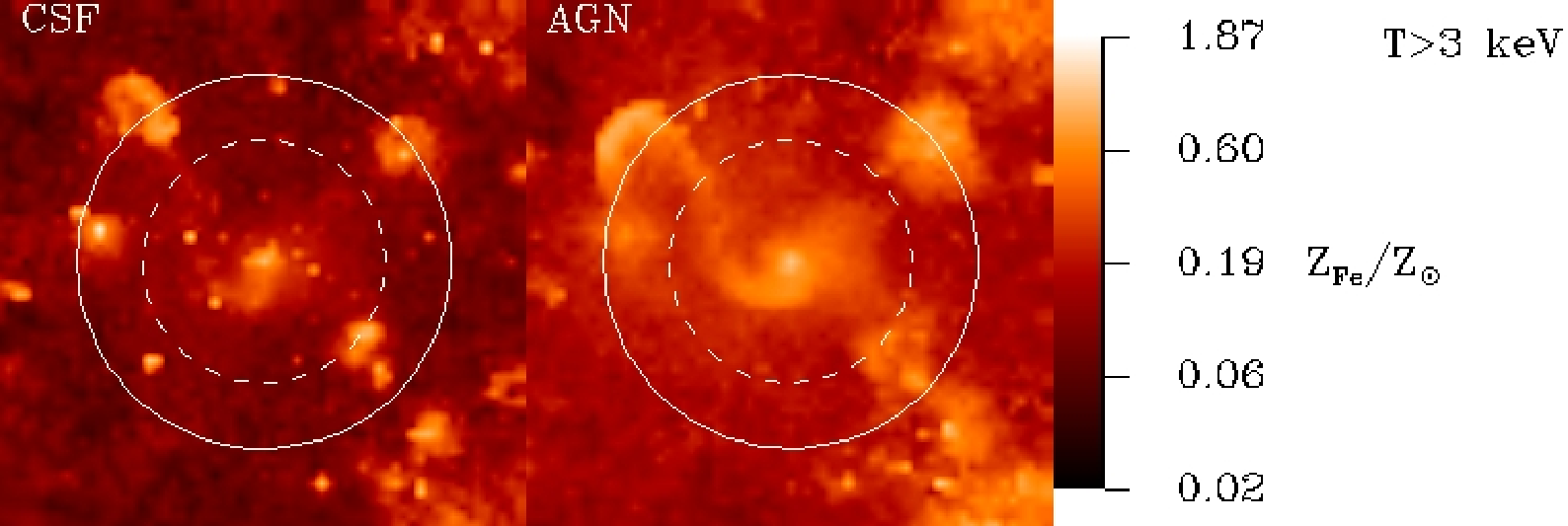}}\\
{\includegraphics[width=15.5cm]{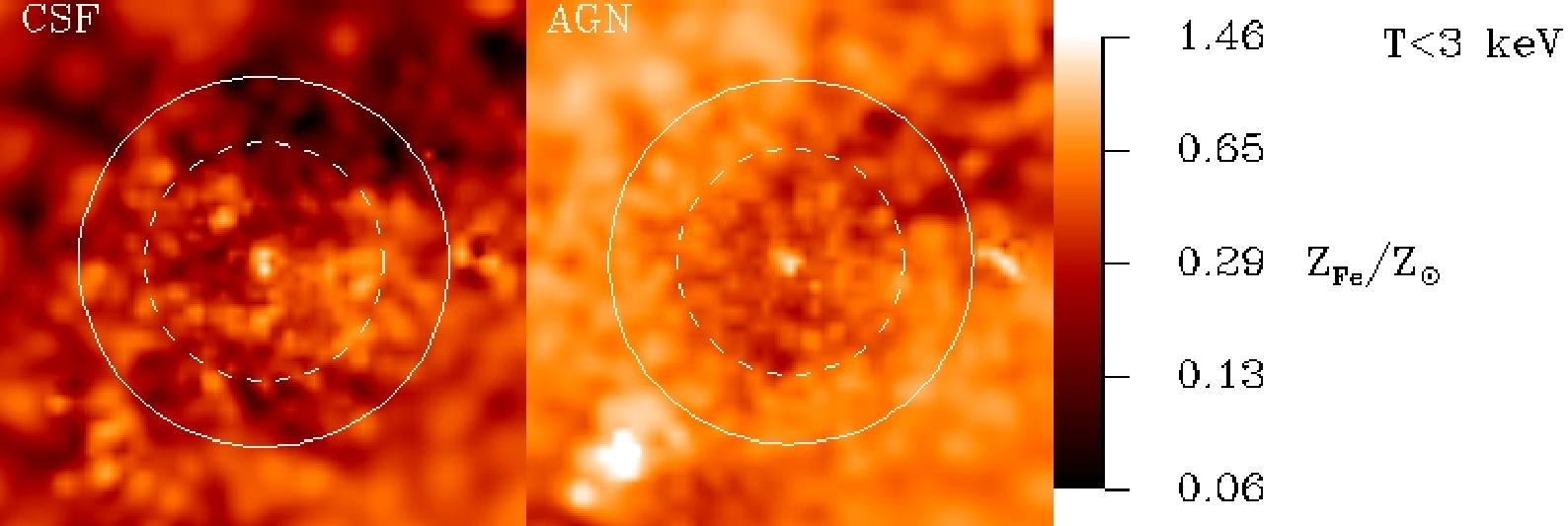}}
\end{center}
\caption{Maps of emission-weighted iron abundance for two different
  systems in the {\tt \w} (left column) and {\tt \agn} (right column)
 simulation sets, respectively.  In the top row the maps are shown for a galaxy
  cluster with $M_{500}\sim 9\times10^{14}\msun$, $T_{sl,500}\sim 7
  keV$ and $R_{vir}\sim 2.5 \hMpc$, whereas in the bottom row a group
  with $M_{500}\sim 3\times10^{13}\msun$, $T_{sl,500}\sim 1 keV$ and
  $R_{vir}\sim 0.85 \hMpc$ is represented.  Abundance values are
  expressed in units of the solar values, as reported by
  \citet{Grevesse1998}, with colour coding specified in the right bar.
  For each object, each map has a side of $2\times R_{vir}$.  White
  circles on each panel represent the values of $R_{180}$ (continuous
  line) and $R_{500}$ (dashed line) for both systems.  We have
  normalized the values of Fe abundance to a common maximum and
  minimum values in order to make comparable the different runs for
  the same system.}
\label{fig:iron_maps}
\end{figure*}

In order to better understand the shape of the emission-weighted Fe
profiles at large radii, we show in Fig.~\ref{fig:iron_maps} the
maps of emission-weighted iron abundance for two different systems, in
the {\tt \w} (left column) and {\tt \agn} (right column) sets,
respectively.  These systems correspond to a galaxy cluster with
$T_{sl,500}\sim 7 keV$ (top row) and a smaller system with
$T_{sl,500}\sim 1 keV$ (bottom row).  Each map has a side of length $2\times
R_{vir}$ and, therefore, they represent a region of $\sim 5 \hMpc$ for
the big galaxy cluster and $\sim 1.6 \hMpc$ for the smaller one.  We
have also highlighted with white circles the values of $R_{180}$
(solid line) and of $R_{500}$ (dashed line) for both systems.

From a visual inspection of these maps we qualitatively appreciate the
effect that different feedback mechanisms have on the ICM enrichment
pattern.  For the more massive system (top row), the distribution of
Fe is quite different in the {\tt \w} and {\tt \agn} cases. Despite
the high efficiency that galactic winds have in spreading metals in
the intergalactic medium at high redshifts \citep[$z\gtrsim2$;
e.g.,][]{Oppenheimer_2008, Tornatore_2010, fabjan_etal10}, winds in
the {\tt \w} case are not efficient enough to regulate star formation
inside clusters. As a result, within $\sim 0.5 R_{180}$, a central
high-metallicity peak is surrounded by a relatively low-metallicity
gas, partially stripped by merging galaxies. In the outer cluster
regions highly-enriched gas clumps are still present.  As explained
above, these clumps are responsible for the increase of the
emission-weighted iron abundance profiles at large radii, shown
in Fig.~\ref{fig:metals_prof_ew}.  From the top-right panel of
Fig.~\ref{fig:iron_maps} we infer that AGN feedback has higher 
efficiency in mixing and distributing the metals through the ICM.  As
a result, instead of a clumpy distribution of highly-enriched gas, we
obtain a rather high level of diffuse enrichment from the centre to
the outskirts.  Within the viral radius of the cluster, the transition
between the enrichment of the central Fe peak and the surroundings is
more continuous than in the {\tt \w} case, thus giving rise to
shallower profiles. For outer cluster regions, although
highly-enriched regions in the outskirts are more diluted than in the
previous case, they contribute in the same way to the slope of the
iron abundance profiles.

The situation is completely different for the smaller system (bottom
row). In this case the level of enrichment of the ICM decreases from
the highly-enriched central core out to the outermost regions in both
radiative runs, although the values of $Z_{Fe}$ are slightly higher in
the case of the {\tt \agn} simulation throughout the cluster volume.
In addition, given the efficiency of AGN feedback in spreading metals
from star-forming regions, the iron distribution is much more uniform
in this case, thus producing flatter $Z_{Fe}$ profiles in outer
regions \citep[see also][]{fabjan_etal10}.

The sensitivity of the metal distribution in cluster  outskirts
on the nature of feedback demonstrates the relevance of pushing
observational analyses of the ICM enrichment out to large
radii. While carrying out such measurements well outside $R_{500}$ is
beyond the reach of the current generation of X-ray telescopes, the
future generation of high-sensitivity instruments, with both large
collecting area and high spectral and angular resolution, would be able to
trace the pattern of ICM chemical enrichment out to such radii, thereby
constraining the past history of cosmic feedback.

\subsubsection{The $Z_{Si}/Z_{Fe}$ relative abundance}

\begin{figure}
\begin{center}
{\includegraphics[width=8.5cm]{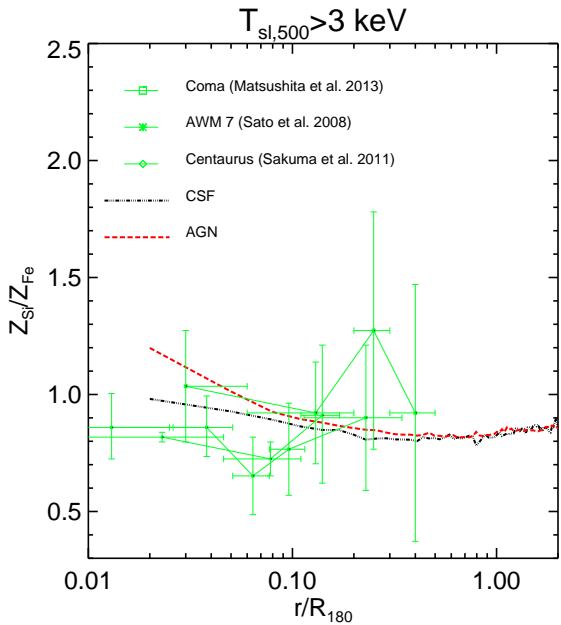}}
\end{center}
\caption{Mean emission-weighted \Zsife profiles obtained by averaging
  over simulated clusters with $T_{sl,500}\ge3$ keV. Black
  dotted-dashed and red dashed lines stand for the {\tt \w} and {\tt
    \agn} runs, respectively.  In order to compare with observational
  results, we use profiles from different \suzaku\ observations. }
\label{fig:sife_prof}
\end{figure}

SNe-Ia produce a large amount of Fe and Ni elements, while SNe-II are
the main contributors of O, Ne, and Mg.  Si-group elements (Si, S, Ar,
and Ca) are produced by both SN types in similar proportions.
Therefore, studying the relative abundance of different elements
potentially offers a means to infer the relative contribution of
different stellar populations to the ICM enrichment and, subsequently, to
reconstruct the stellar IMF \citep[e.g.][and references
therein]{Loewenstein_2013}.

Early ASCA data \citep[e.g.,][]{Loewenstein_1996, Fukazawa_1998,
Finoguenov_2000} suggested that cluster outskirts are
predominantly enriched by SNe-II. A similar result was found more
recently by \cite{Rasmussen_2007} and \cite{Rasmussen_2009}. These
authors analysed \xmm\ data for poor clusters with $T\mincir 3$ keV
obtaining rather flat profiles of \Zsife with a value close to solar
at small radii, followed by a steep rise beyond $\sim0.2 R_{500}$.
\suzaku\ observations of low temperature clusters and groups
\citep[e.g.,][and references therein]{Sato_2010, Sakuma_2011, Matsushita_2013}   
show instead rather flat profiles of $Z_{Si}/Z_{Fe}$ out to large
radii, $\simeq 0.3 R_{vir}$, thus implying that SNe-Ia and SNe-II
should contribute in similar proportions to the enrichment at
different radii.

Figure \ref{fig:sife_prof} shows the mean emission-weighted \Zsife
radial profiles for our sample of high-temperature systems. As in
previous simulations \citep[e.g.,][]{fabjan_etal10, McCarthy_2010}, we
find that Si/Fe abundance ratio is nearly flat as a function of radius
from roughly $\sim 0.1 R_{180}$ out to the outermost radius.  Our
results are qualitatively in good agreement with different \suzaku\
observations \citep[e.g.,][]{Sato_2008, Sakuma_2011, Matsushita_2013},
although within fairly large uncertainties.  Simulations with AGN
feedback produce a relative increase of the Si abundance in the
central regions \citep[see also][]{fabjan_etal10}.  This increase is due
to the selective removal of metal-enriched gas associated  with radiative
cooling.  Total metallicity of the gas around the BCG is dominated by
SNe-II products. Therefore, gas more enriched by SNe-II has a
relatively shorter cooling time. As a consequence, suppression of
cooling in the core regions by AGN feedback tends to increase the
amount of SN-II products in the ICM, thereby justifying the increase
of \Zsife with respect to the {\tt \w} runs.

A central relative enhancement of silicon abundance, as found in the
{\tt \agn} case, is marginally disfavoured by available
observations. This may hint at a lack of some diffusion or
transport process in our simulations, which are responsible for mixing
different metal species. In our simulations, the effect of AGN is to
provide a purely thermal feedback. In this respect, the explicit
inclusion of kinetic feedback in the form of sub-relativistic jets is
expected to trigger turbulence and circulation of gas which may have a
significant effect on the distribution of metals in central cluster
regions.

\subsection{Metallicity of stars}
 
\begin{figure}
\begin{center}
{\includegraphics[width=8.5cm]{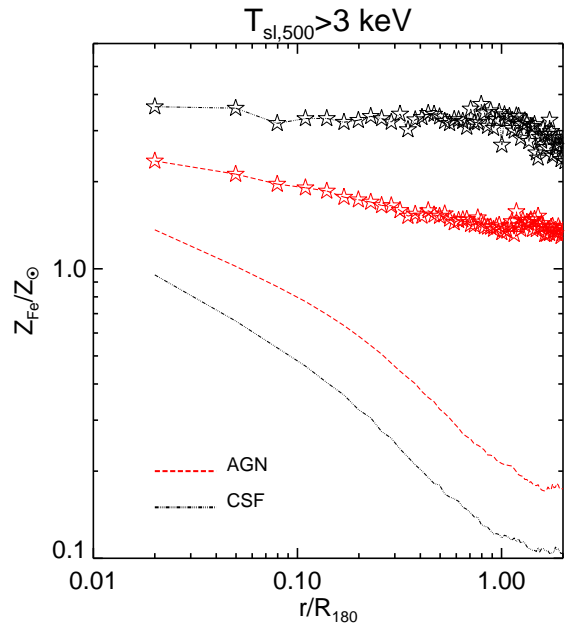}}
\end{center}
\caption{Mean mass-weighted iron abundance profiles for the gas and
  the stellar components in our {\tt \w} and {\tt \agn} runs. Black
  dotted-dashed and red dashed lines represent the $Z_{Fe}$ in gas of
  the {\tt \w} and {\tt \agn} runs, respectively. The same type of lines
  connected by star symbols represent the corresponding $Z_{Fe}$ in
  the stellar component for each run. These profiles have been
  obtained by averaging over the sample of hot systems with
  $T_{sl,500}\ge3 keV$ out to $2\times R_{180}$.}
\label{fig:stars_metals_prof_ew}
\end{figure}

Given the significant suppression of star formation in the simulations
including AGN feedback, it may appear surprising that these
simulations are characterized by a higher level of ICM enrichment.
In fact, the explanation for this behaviour lies in the different
efficiency with which metals, after being distributed from star
particles to the surrounding gas particles, are later locked back into
stars.

Figure \ref{fig:stars_metals_prof_ew} compares the mean mass-weighted
iron abundance profiles (already displayed in
Fig.~\ref{fig:metals_prof_ew}) together with the corresponding mean
profiles for the iron abundance in stars.  A comparison between the
two simulations shows nearly an order of magnitude difference in the
normalization of the mean profiles associated with the stellar
component, with the {\tt \w} run being much more metal rich than those
in the {\tt \agn} run.  The effect of feedback from SMBHs is therefore
not only to quench the overall star formation rate, but also to
prevent highly-enriched gas particles, which have a short cooling
time, from undergoing rapid star formation.  BHs heat the surrounding
gas causing it to expand out of the dense, star-forming regions,
thereby raising the fraction of the total metal mass in the gas phase.
The efficient ejection of metals from star-forming regions is the
mechanism by which the {\tt \agn} simulations yield ICM metal
abundances similar to, or even larger than, those in the {\tt \w} case, in
spite of a stellar mass fraction that is $\sim 3$ times smaller
\citep[see also][]{sijacki_etal07, fabjan_etal10, McCarthy_2010, Planelles_2012}.

These results highlight that the nature of feedback affects not only
the star formation within cluster galaxies, but also the way in
which metals are distributed between the hot ICM and
stars. Independent analyses based on semi-analytical models of galaxy
formation \citep[e.g.,][]{DeLucia_2012, Henriques_2013} have in fact
demonstrated that the metal content of galaxies, and most prominently
of the BCGs, are inextricably linked to the past history of star
formation and provide quite stringent constraints on the nature of
feedback.  A complete analysis of the properties of the galaxy
population in simulations will be presented in a future analysis.

\section{Summary and discussion}
 
We have analysed a set of cosmological hydrodynamical simulations of
galaxy clusters paying special attention to the effects that different
implementations of baryonic physics have on the thermal and
chemodynamical properties of these systems. Using the Tree--PM SPH
code {\tt GADGET-3} \citep{springel05}, we carried out re-simulations
of 29 Lagrangian regions extracted around as many galaxy clusters
identified within a low-resolution N-body parent simulation. These
cluster re-simulations have been performed using different
prescriptions for the baryonic physics: without including any
radiative processes ({\tt \nr} runs), including the effect of cooling,
star formation and SN feedback ({\tt \w} runs), and including also an
additional contribution from an efficient model of AGN feedback ({\tt
  \agn} runs).

The final sample of objects obtained within each one of these sets of
re-simulations consists of $\simeq 160$ galaxy clusters and groups
with $M_{vir}\ge3\times 10^{13}\msun$ at $z=0$.  Using these three
sets of simulated galaxy clusters, we have analysed how star formation
and feedback in energy and metals from SN and AGN affect their X-ray
scaling relations and associated radial profiles, and the chemical
enrichment pattern of their hot intra-cluster gas.  Our main results
can be summarised as follows.

\begin{itemize}

\item Including gas accretion onto SMBH and the ensuing AGN feedback
  provides, in general, a better agreement between simulation results
  and observations of X-ray scaling relations. The differential
  effect that AGN feedback has at the scales of groups and of massive
  clusters is such to change the slope and normalization of the
  mass--temperature relation and bring it into better agreement with
  observational results with respect to the {\tt \nr} and {\tt \w}
  cases. In a similar way, injection of entropy from AGN feedback
  causes a suppression of gas density at the centre of groups,
  relative to clusters, thereby bringing the \lxtsl\ and \stsl\ relations into
  closer agreement with observations \citep[see also][]{Puchwein2008,
  short_2010, fabjan_etal10, McCarthy_2010}.

\item Feedback also has an impact on the thermal structure of the ICM and,
  in turn, on the difference between mass-weighted and
  spectroscopic-like temperature, the latter being sensitive to the
  presence of clumps of relatively cold gas associated with substructures,
  from which it is eventually ram--pressure stripped. Removal of cold
  gas in substructures by radiative cooling or by the effect of
  efficient AGN feedback goes in the direction of decreasing the
  thermal complexity of the ICM and, therefore, the difference between
  \tmw\ and \tsl. A careful comparison with observations on the
  temperature distribution of the ICM \citep[e.g.][]{Frank_2013} is required in
  order to verify whether our simulations provide the correct degree
  of thermal complexity in clusters.
  
\item Simulations including AGN feedback recover both
  X-ray and SZ results on the pressure profiles quite well. Although we also
  find a reasonable agreement with observations of temperature and
  entropy profiles, our simulations do not produce the observed
  difference in profiles between relaxed, cool-core clusters and
  unrelaxed, non cool-core systems, even when AGN feedback is
  included. We regard this as a major limitation of the implementation
  of AGN feedback included in our simulations. Ultimately, this traces
  back to the limited capability of current implementations of AGN
  feedback in cosmological simulations to produce the correct
  structure of cool cores.
    
\item In broad agreement with observations \citep[e.g.,][]{Zhang2011},
  we obtain a decreasing iron abundance with increasing total cluster
  mass. However, observational data reveals a decrease of iron
  abundance in high-mass systems that is stronger than predicted by
  our simulations. Such a difference is even worsened by the inclusion
  of AGN feedback. In this case, highly--enriched gas in massive
  clusters is prevented from cooling out of the hot phase, thereby
  increasing the enrichment level with respect to simulations
  including only SN feedback. This tension is caused by the still
  insufficient regulation of star formation provided by simulated AGN
  feedback at the centre of massive
  halos \citep[e.g.][]{Puchwein_2013}, which also causes simulated
  BCGs to be too massive with respect to the observed ones
  \citep[for a detailed analysis see][and references therein]{RagoneFigueroa_2013}.

\item Despite the exceedingly high level of iron abundance predicted
  in massive clusters, the shape of the radial profiles of Fe
  abundance is in better agreement with the observed profiles when AGN
  feedback is included. In general, we find that AGN produces a higher
  and much more widespread pattern of metal enrichment in the outer
  part of galaxy clusters and groups. In fact, AGN feedback, combined
  with galactic winds, is quite efficient in ejecting highly-enriched
  gas from star-forming regions, thus enhancing the metal circulation
  in the inter-galactic medium. This generates a rather uniform and
  widespread pattern of metal enrichment in the outskirts of clusters.

\item Our simulations predict almost flat profiles of \Zsife
  out to the outermost radii. Suppression of star formation in the
  runs with AGN feedback causes \Zsife to increase at small radii,
  $\mincir 0.1R_{500}$, a feature which is possibly in tension with
  the rather uncertain observational results. If confirmed, this
  tension may suggest that some additional mechanisms, not included in our
  simulations, are responsible for efficient mixing of metals in
  the central cluster regions.
 
\item Despite the suppression of star formation in the {\tt \agn}
  simulations, they produce a higher level of enrichment in massive
  clusters. The reason for this apparently paradoxical result is that
  AGN feedback is efficient in preventing highly--enriched gas to leave
  the hot phase and form stars. As a result, we consistently find stellar
  metallicity in the {\tt \agn} simulations to be suppressed by almost
  a factor of 2 with respect to the {\tt \w} case.
    
\end{itemize} 

Our results support the idea that including feedback from
SMBHs significantly improves the ability of cosmological
hydrodynamical simulations to yield a realistic population of galaxy
clusters and groups. Indeed, this result agrees qualitatively with previous works
that implemented BH growth and feedback in cosmological 
simulations of galaxy clusters and groups \citep[e.g.,][]{sijacki_etal07, Puchwein2008,
 fabjan_etal10, McCarthy_2010}. These findings are quite encouraging,
especially if we keep in mind that relatively simple prescriptions are
adopted to describe the rate of gas
accretion and the thermalisation  of the extracted energy. An
interesting prediction of these simulations is that the pattern of
metal distribution in the cluster outskirts represent a fossil
record of the interplay between feedback and star formation during the 
hierarchical assembly of clusters. While tracing the enrichment of the
inter-galactic medium in this regime is beyond the capabilities of the
available X-ray telescopes, future instruments are expected to have
the required sensitivity and spectroscopic capability, thereby
shedding light on a crucial aspect of galaxy evolution. 

Despite the above successes, our results also highlight that a number of discrepancies
between observations and predictions from simulations still
exist. Even including AGN feedback we are not able to produce the
correct cooling/heating interplay in cluster cores. This limitation
manifests itself in, e.g., BCGs that are too large and the lack of diversity 
of ICM properties between relaxed and unrelaxed systems.

One aspect of our AGN implementation that needs to be improved is
related to the pure thermal nature of the associated
feedback. Observational results indicate that sub-relativistic jets
from the BH hosted in central cluster galaxies shocks the
surrounding ICM, thereby producing bubbles of high-entropy gas. 
Controlled numerical experiments of isolated clusters
\citep[e.g.][]{Omma_2004,Brighenti_2006,Gaspari_2011} and disk
galaxies, or controlled galaxy mergers
\citep[][]{Choi_2012,Barai_2013} have already demonstrated that
kinetic AGN feedback provides results that are rather different from
those based on thermal feedback. In particular, outflows generate
circulation of gas that has the twofold effect of stabilising cooling
flows and distributing enriched gas outside the innermost regions.
Including mechanical AGN feedback in cosmological simulations is
clearly a step to be undertaken \citep[e.g.][]{Martizzi_2012}, along
with the exploration of accretion models different from the standard
Bondi criterion.

\section*{ACKNOWLEDGEMENTS} 
 
The authors would like to thank Volker Springel for making available
to us the non-public version of the {\small GADGET--3} code and
Annalisa Bonafede for her help with generating the initial conditions
for the simulations. We also would like to thank Gabriel Pratt,
Dominique Eckert and Kyoko Matsushita for supplying us with
observational data to compare with, and the anonymous referee for
his/her constructive comments that helped improving the presentation
of the results.  Simulations have been carried out at the CINECA
supercomputing Centre in Bologna, with CPU time assigned through ISCRA
proposals and through an agreement with University of Trieste.  SP and
MK acknowledge a fellowship from the European Commission's Framework
Programme 7, through the Marie Curie Initial Training Network
CosmoComp (PITN-GA-2009-238356).  DF acknowledges founding from the
Centre of Excellence for Space Sciences and Technologies SPACE-SI, an
operation partly financed by the European Union, European Regional
Development Fund and Republic of Slovenia, Ministry of Higher
Education, Science and Technology.  CR-F acknowledges founding from
the Consejo Nacional de Investigaciones Cient\'{\i}ficas y T\'ecnicas
de la Rep\'ublica Argentina (CONICET) and the Secretar\'{\i}a de
Ciencia y T\'ecnica de la Universidad Nacional de C\'ordoba -
Argentina (SeCyT).  This work has been supported by the PRIN-INAF09
project ``Towards an Italian Network for Computational Cosmology'', by
the PRIN-MIUR09 ``Tracing the growth of structures in the Universe'',
and by the PD51 INFN grant.  This work has been partially supported by
{\it Spanish Ministerio de Ciencia e Innovaci\'on} (MICINN) (grants
AYA2010-21322-C03-02 and CONSOLIDER2007-00050.

\bibliographystyle{mnbst}
\bibliography{dianoga_paper}

\end{document}